\UseRawInputEncoding
\documentclass[10pt,twocolumn]{IEEEtran}

\usepackage{mathrsfs}
\usepackage{pifont}
\usepackage{bbding}
\usepackage{tipa}
\usepackage{color}
\usepackage{cite}
\usepackage{graphicx}
\usepackage{url}
\usepackage{amsfonts}
\usepackage{float}
\usepackage{times}
\usepackage{psfrag}
\usepackage{subfigure}
\usepackage{stfloats}
\usepackage{footnote}
\usepackage{array}
\usepackage{amsmath}
\usepackage{stmaryrd}
\usepackage{amssymb}
\usepackage{epic}
\usepackage{graphicx}
\usepackage{curves}
\usepackage{balance}
\usepackage{bm}
\usepackage{tabularx}
\usepackage{xcolor}
\usepackage{graphics}
\usepackage{epsfig}
\usepackage{epstopdf}
\usepackage{enumerate}
\usepackage{color}
\usepackage{lettrine}
\usepackage{mdwlist}
\usepackage{threeparttable}
\usepackage[linesnumbered,ruled,vlined]{algorithm2e}
\usepackage{algpseudocode}
\usepackage{lipsum}
\usepackage{makecell}
\usepackage{ntheorem}

\ifCLASSINFOpdf
\else
\fi

\begin{document}
\bstctlcite{IEEEexample:BSTcontrol}
\theoremheaderfont{\bfseries\upshape}
\theoremseparator{:}
\newtheorem{theorem}{Theorem}
\newtheorem{definition}{Definition}
\newtheorem{problem}{Problem}
\newtheorem{lemma}{Lemma}
\newcommand{\upcitep}[1]{\textsuperscript{\textsuperscript{\citep{#1}}}}

\title{Federated Learning Over Wireless Channels: Dynamic Resource Allocation and Task Scheduling}

\author{Shunfeng~Chu,
        Jun~Li,
        Jianxin~Wang,
        Zhe~Wang,
        Ming~Ding,
        Yijin~Zang,
        Yuwen~Qian,
        and~Wen~Chen
\thanks{S. Chu, J. Li, J. Wang, Y. Zhang and Y. Qian are with the School of Electronic and Optical
Engineering, Nanjing University of Science and Technology, Nanjing
210094, China (e-mail: shunfeng.chu@njust.edu.cn; jun.li@njust.edu.cn; wangjxin@njust.edu.cn; yijin.zhang@njust.edu.cn; admon@njust.edu.cn).}
\thanks{Z. Wang is with the School of Computer Science and Engineering, Nanjing University of Science and Technology, Nanjing 210094, China (e-mail: zwang@njust.edu.cn).}
\thanks{M. Ding is with Data61, CSIRO, Sydney, Australia (e-mail: ming.ding@data61.csiro.au).}
\thanks{W. Chen is with the Department of Electronic Engineering, Shanghai Jiao Tong University, Shanghai 200240, China (e-mail: wenchen@sjtu.edu.cn}}

\markboth{submitted to IEEE Transactions On Network And Service Management}%
{Shell \MakeLowercase{\textit{et al.}}: Bare Demo of IEEEtran.cls for IEEE Journals}
%



\maketitle

\begin{abstract}
With the development of federated learning (FL), mobile devices (MDs) are able to train their local models with private data and sends them to a central server for aggregation, thereby preventing sensitive raw data leakage. In this paper, we aim to improve the training performance of FL systems in the context of wireless channels and stochastic energy arrivals of MDs.
To this purpose, we dynamically optimize MDs' transmission power and training task scheduling. We first model this dynamic programming problem as a constrained Markov decision process (CMDP).
Due to high dimensions rooted from our CMDP problem, we propose online stochastic learning methods to simplify the CMDP and design online algorithms to obtain an efficient policy for all MDs. Since there are long-term constraints in our CMDP, we utilize Lagrange multipliers approach to tackle this issue. Furthermore, we prove the convergence of the proposed online stochastic learning algorithm. Numerical results indicate that the proposed algorithms can achieve better performance than the benchmark algorithms.
\end{abstract}


\IEEEpeerreviewmaketitle

\section{Introduction}
%
%
%
%
\IEEEPARstart{I}{n} the last decade, we have witnessed a series of amazing breakthroughs, such as AlphaGo, machine learning and artificial intelligence (AI), which have become the most cutting-edge technology in both academia and industry communities \cite{lecun2015deep}.
Distributed machine learning based on mobile edge computing (MEC) of the wireless networks is also one of the current hot research directions \cite{mao2018deep}. The sample data for machine learning can be obtained by collecting massive amounts of data from mobile devices (MD) distributed in the wireless network.
By training these data, the training performance of machine learning can be greatly improved.

Although offloading the local sample data of distributed MDs for centralized learning significantly improves the performance of machine learning, this mechanism suffers from two flaws. First, transmission delays from distributed MDs to the central cloudy via backbone network are extremely large. Second, the local data often contains the private information of MDs, and uploading the private information to the central cloudy will lead to the risk of personal privacy leakage. To cope with these two issues, federated learning (FL) has been introduced to act as an emerging distributed machine learning paradigm, which is indeed a way to combine MEC with traditional machine learning. In this manner, MDs train their local data and send their local model updates to a task publisher iteratively instead of uploading the raw data to a central server \cite{wei2020federated,wei2021user}, which, in general, brings the following two benefits. (\uppercase\expandafter{\romannumeral1}) The communication latency and the energy consumption for computation can be significantly reduced owing to the fact that MDs are not required to upload huge amounts of local data for training to an edge server. (\uppercase\expandafter{\romannumeral2}) MDs upload their local model instead of the raw data to the edge server, which greatly reduces the risk of personal privacy information leakage \cite{ma2020on}.

Despite the aforementioned advantages of FL, there are still many challenges that have not been solved until now. Some existing studies \cite{wang2019adaptive,kim2019blockchained} adopted an idealized assumption that all MDs participating in FL are immune to the wireless and computation resource constraints. \cite{mcmahan2016communication} only focused on a practical Federated-Averaging algorithm for distributed DNN training and the training performance. Many studies \cite{li2014communication,alistarh2017qsgd,lin2017deep,sattler2019sparse} have been committed to further reducing the communication overhead by developing compression methods.
However in practice, MDs usually suffer from energy consumption constraints that may reduce the network lifetime and training efficiency.

In addition, frequent wireless communications is usually required for uploading and downloading the model parameters, which would increase the bandwidth cost and the training latency \cite{anh2019efficient}. Therefore, it is necessary to design a feasible resource scheduling scheme to optimize the resource scheduling problem in the FL process. Without taking into account the energy constraints and battery dynamics
of MDs, \cite{tran2019federated} formulated a FL over a wireless networks as a static optimization problem, and exploited the problem structure to decompose it into three static convex sub-problems. The work \cite{mohammad2019adaptive} proposed a static scheduling scheme to efficiently execute distributed learning tasks in an asynchronous manner while minimizing the gradient staleness on wireless edge nodes with heterogeneous computing and communication capacities. However, many properties of MEC networks are usually time-varying in the FL process and thus the methods in \cite{mao2016grid} will result in considerable performance loss. The work \cite{mao2016grid,mao2015a} model the channel and energy dynamically, and exploit the dynamic scheduling algorithm to obtain the asymptotically optimal resluts. Thus, it is of vital importance to develop efficient dynamic resource scheduling schemes to improve the performance of FL.

In this paper, we utilize constrained Markov decision processes (CMDP) as a mathematical tool to obtain an optimal algorithm for dynamic resource scheduling for FL, where each MD send local model updates
trained on their local raw data iteratively to a common edge server, and the edge server aggregates the parameters from MDs participating in local training and broadcasts the aggregated parameters to all the MDs. In particular, each MD possesses computing units with computing capability, which can be used for local machine learning with local raw data. In order to improve the training performance of FL\footnote{In general FL, existing work \cite{yang2020FL} used the accuracy of the test set to measure the performance of machine learning after training. Since the quantitative analysis of the accuracy in the test set is relatively difficult, we use the size of local dataset accumulated from MDs over iterations to evaluate the accuracy of the machine learning model \cite{anh2019efficient}.}, we propose an efficient stochastic optimization algorithm for scheduling resources of MDs in the FL processes by obtaining an efficient setting of the size of raw data for local training, and the transmission power of MDs to upload the local model.

Our main contributions are listed as follows.
\begin{enumerate}
  \item
  Due to the dynamic nature of wireless network and battery status of MDs, we consider resource scheduling of FL in dynamic scenarios.
  Thus, we model the resource scheduling problem of the FL process as a CMDP problem, and improve the performance of FL
  by optimizing the size of the local training data at the MD side.
  \item Since the state-action space dimension in the constrained MDP problem is large and there are a few constraints in the dynamic problem,
  we simplify the stochastic optimization problem by proving an equivalent Bellman equation and using the Lagrange multiplier method.
  \item We use approximate MDP and stochastic learning methods to analyze the constrained MDP problem, and design centralized online algorithms to obtain
  resource scheduling policy for all MDs.
  \item We also provide effective analysis for the convergence of the online stochastic learning algorithms.
\end{enumerate}

Although the idea of applying CMDP to design dynamic resource scheduling is not new, we are motivated to address the resource scheduling issues of resource constraints and dynamics of FL. To achieve high-quality learning performance,
a reasonable constrained dynamic scene is essential to the resource scheduling issues \cite{wu2021multi,wu2019dynamic}. Previous work has utilized CMDP as a mathematical model to design effective algorithms for resource scheduling in wireless networks \cite{anh2019efficient,van2018quality,zhao2019delay}, which is considered as an effective tool for solving dynamic problems and time-related state problems. Inspired by this, we apply CMDP as the mathematical scene to address the resource scheduling problem in the FL process. Nevertheless, previous work still has some shortcomings in solving CMDP problems. The literature \cite{anh2019efficient} adopted a deep learning algorithm that allows the edge server to learn and find optimal decisions without any a priori knowledge of network dynamics in the CMDP. However, reinforcement learning is poorly scalable and requires a lot of computing power and duration for training. The literature \cite{van2018quality} obtained the optimal solution for the formulated CMDP offloading problem by linear programming and Q-learning method. Neither method is suitable for large-scale networks, which will lead to the curse of dimension. The work \cite{zhao2019delay} developed a threshold-based algorithm to obtain the optimal delay-power tradeoff efficiently, in which the authors used the special structure of the mathematical model to solve the CMDP problem, and this method does not possess the universality of solving the problem.

The rest of this paper is organized as follows. Section \ref{sec:System Model} describes the system model and dynamic analysis. CMDP-based dynamic resource scheduling problem is formulated in Section \ref{Constrained Markov Decision Problem Formulation}. Section \ref{approximate_Markov_decision process1} proposes approximate Markov decision process and stochastic learning methods to simplify the CMDP problem, and designs online algorithms to obtain an efficient policy. Section \ref{sumulation1} presents the simulation results. Finally, Section \ref{sec:conclusion} concludes this paper.

\setlength{\tabcolsep}{1mm}{
\begin{table}
\caption{Summary of main notations}
\centering
\begin{tabular}{l||l}
\hline
\textbf{Notation}& \textbf{Description}\\
\hline\hline
$N$& The number of all MDs\\
\hline
$n$& The index of the MD\\
\hline
$t$& The number of iterations of FL\\
\hline
$b_n(\cdot)$& The number of bits of training data for the $n$-th MD\\
\hline
$P_n(\cdot)$& The transmission power of the $n$-th MD \\
\hline
$E_n^{\text{max}}$& The battery capacity of the $n$-th MD\\
\hline
$E_n^{\text{sta}}(\cdot)$& The $n$-th MD's energy state\\
\hline
$E_n^{\text{cop}}(\cdot)$& The $n$-th MD's computation energy for local model training\\
\hline
$E_n^{\text{com}}(\cdot)$& \makecell[l]{The $n$-th communication energy for parameter transmission}\\
\hline
$E_n^{\text{arr}}(\cdot)$& The harvesting energy for the $n$-th MD\\
\hline
$C_n$& The CPU cycles to train a unit sampled data on the $n$-th MD\\
\hline
$f_n$& The CPU frequency of the $n$-th MD\\
\hline
$\tau_n^{\text{cop}}(\cdot)$& The processing time of local training on the $n$-th MD\\
\hline
$\epsilon_n(\cdot)$& The upload decision of the $n$-th MD\\
\hline
$h_n(\cdot)$& \makecell[l]{The channel gain between the $n$-th MD and the edge server}\\
\hline
$R_{n}(\cdot)$& The uplink transmission rate for the $n$-th MD\\
\hline
\end{tabular}
\label{tab:summ_nota}
\end{table}

\section{System Model}\label{sec:System Model}
Consider a wireless FL system consisting of an edge server and $N$ MDs as shown in Fig.~\ref{fig:system model figure}. Each MD is equipped with computing and energy harvesting modules. Having access to a vast range of local data, each MD is able to train the machine learning model locally using the harvested energy from the environment. To improve the model training efficiency and protect data privacy, the FL technique is adopted as an iterative model updating process between the edge server and MDs.

We first briefly introduce the main procedures as follows. In each learning iteration $t$, the $n$-th MD first selects $b_n(t)$ bits of training data from the local data set, where the size of selected data is determined by the edge server according to the energy status of the MD, i.e., the battery energy at the beginning of this iteration. Here, we assume that the edge server knows the energy status of all MDs in advance at the beginning of the learning iteration.
Then, each MD performs local training and obtains the local model parameters. Afterwards, the edge server decides whether or not to upload the local parameters for MDs for aggregation according to MDs' remaining energy and channel state. If the $n$-th MD participates, it transmits the parameters to the edge server in the uplink using the power of $P_n(t)$. Finally, the edge server aggregates the local training parameters from all the participating MDs and then it broadcasts the updated global training parameters (e.g., weighted average over the local parameters) back to all the MDs. At the end of this iteration, each MD opportunistically harvests energy from the environment and stores the energy in the rechargeable battery. The above process is repeated until the learning model reaches the desired accuracy level.

In the following subsections, we will explain the above process in more details. We mainly discuss each learning iteration in three stages: dynamic energy harvesting, local model training, model parameter transmission and aggregation.

\begin{figure}
\centering
\includegraphics[width=0.47\textwidth]{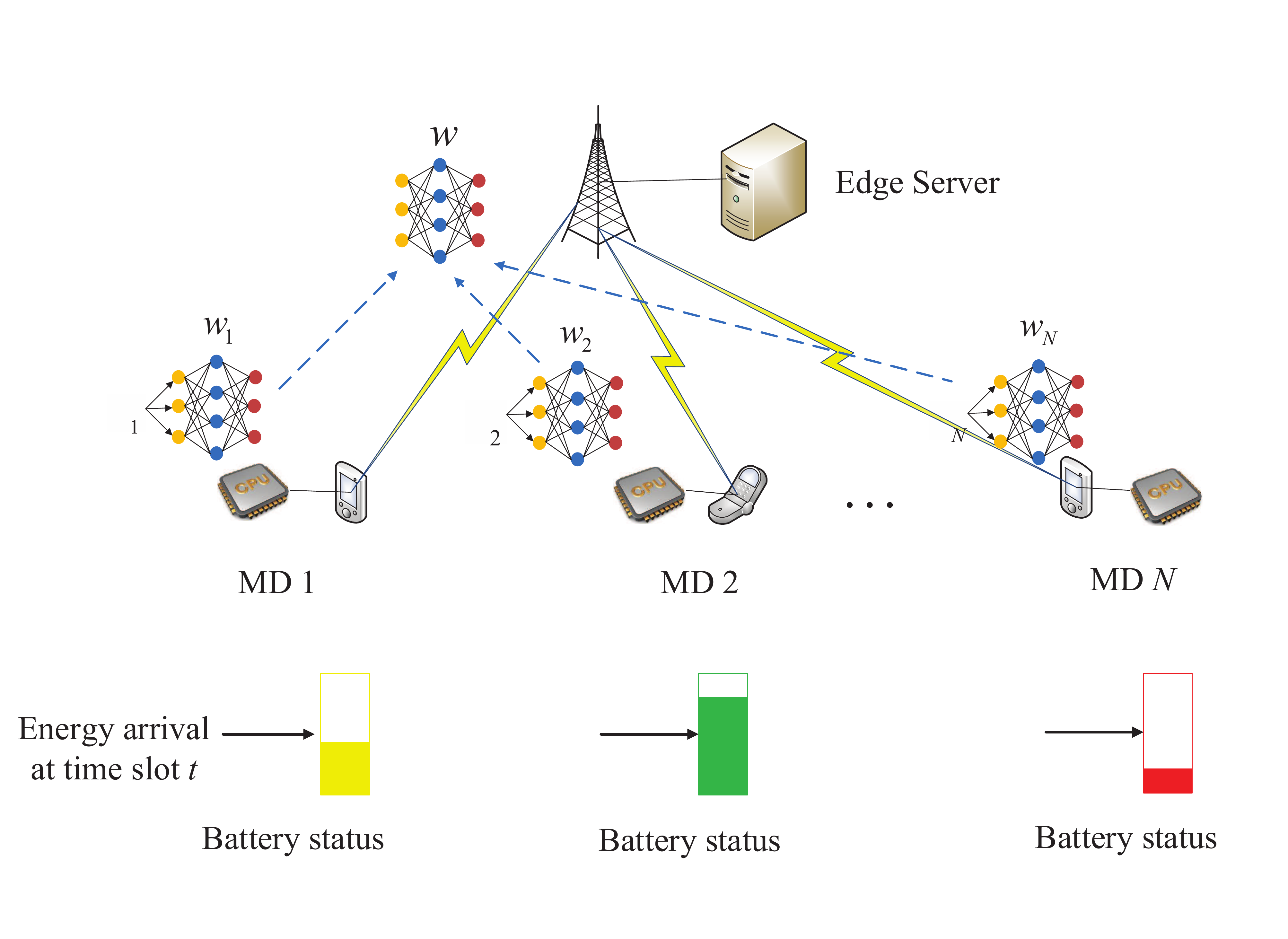}
\caption{The federated learning aided MEC network}
\label{fig:system model figure}
\end{figure}

\subsection{Dynamic Energy Harvesting}
We assume that the $n$-th MD is equipped with a rechargeable battery with a limited capacity of $E_n^{\text{max}}$.\footnote{For the ease of analysis, we quantize the battery capacity in to $E_n^{\text{max}}+1$ uniform levels $\{0,1,\cdots,E_n^{\text{max}}\}$ \cite{Biason2018adecentralized}.} At the beginning  of each iteration $t$, we denote the $n$-th MD's energy state as $E_n^{\text{sta}}(t)$ which is the remaining energy carried from the previous iteration.  According its energy state, the edge server decides whether or not to proceed to local model training and uplink parameter transmission for each MD. We let $E_n^{\text{cop}}(t)$ denote the $n$-th MD's computation energy for local model training  and $E_n^{\text{com}}(t)$ denote the communication energy for uplink parameter transmission, respectively, which will be discussed in more details in the next two subsections. Note that each MD's energy consumption cannot exceed the energy state in this iteration. At the end of the iteration $t$, we consider that each MD is able to harvest energy from the environment and store the energy in the rechargeable battery.  We denote the energy harvesting process for the $n$-th MD by $\{\boldsymbol{E}_n^{\text{arr}}(\cdot): n\in \mathcal{N}\}$, which follows an independent stationary Poisson distribution with average arrival rate $\mathbb{E}[E_n^{\text{arr}}]=\lambda_n$ \cite{Knorn2015distortion}.

Similar to \cite{Sharma2020acceleated}, the energy state of the $n$-th MD at the beginning of iteration $t+1$ can be updated by the following recursion,
\begin{equation}\label{state of the battery from the current time slot}
\begin{split}
E_n^{\text{sta}}(t + 1) = \min \Big\{ [ E_n^{\text{sta}}(t)&-\left\lceil E_n^{\text{com}}(t)+ E_n^{\text{cop}}(t) \right\rceil ]^{+}\\
&+ E_n^{\text{arr}}(t),E_n^{\text{max}} \Big\}, \quad t\geq 1,
\end{split}
\end{equation}
where $\lceil \cdot \rceil$ denotes the ceiling operator, and $x^+ \triangleq \max \{x,0\}$.

\subsection{Local Model Training}
At the beginning of each learning iteration $t$, each MD first selects $b_n(t)$ bits of data samples from the local dataset to perform a machine learning algorithm, and then it obtains the local model parameters. Intuitively, the choice of $b_n$ depends on the available energy in the battery. The MD can train a larger size of the training data if it has more sufficient battery energy in this iteration. Otherwise, it trains less data or takes no training for this iteration. We assume it consumes  $C_n$ CPU cycles to train a unit sampled data on the $n$-th MD. The CPU frequency, denoted by $f_n$  (in CPU cycle/s), is considered as a measurement of computation capacity of the $n$-th MD.
In iteration $t$, the processing time of local training on the $n$-th MD is given by
\begin{equation}
\label{t_c_b}
\tau_n^{\text{cop}}(t) = \frac{{b_n(t){C_n}}}{{f_n}},
\end{equation}
According to \cite{tran2019federated}, the computation energy $E_n^{\text{cop}}(t)$
consumed by local training of the $n$-th MD in the iteration \emph{t} is given by
\begin{equation}
\label{f_c alpha}
E_n^{\text{cop}}(t) = \alpha b_n(t){C_n}{f_n^2},
\end{equation}
where $\alpha$ is the effective capacitance of the computing chipset for each MD.

\subsection{Model Parameter Transmission and Aggregation}
After performing the local training, the MDs then upload their updated local model parameters back to the edge server. Let $\epsilon_n(t)$ denote the upload decision of the $n$-th MD at the iteration $t$, where $\epsilon_n(t)=1$ means the $n$-th MD is assigned to a subchannel and is willing to upload parameters to the edge server through the assigned channel, and $\epsilon_n(t)=0$ indicates that it is not assigned to a subchannel or keeps silent.  Intuitively, an MD is more likely to upload if it has sufficient remaining energy while in a good channel state. For the uplink transmission, we adopt OFDMA technique, where the channels are orthogonal cross the different links. We assume that there are $L$ orthogonal subchannels in the FL system and each MD can only occupy at most one subchannel. Let $h_n(t)$ denote the uplink channel gain between the $n$-th MD and the edge server in the iteration \emph{t}, where the channel gains of all sub-channels between the server and a single MD are same.

We model the channel gain $h_n(t)$ as a discrete-state block fading, where the channel gain between the $n$-th MD and edge server stay is discrete random variable with a general distribution $\Pr[\bar{h}_n]$ \cite{cui2010distributive,zhao2020delay,cui2012survey,zhao2019non,ku2015onenergy}. We further assume that $h_n(t)$ stays invariant within each iteration and are independently and identically distributed (i.i.d.) across different iterations and MDs.

If the $n$-th MD is allowed to upload ($\epsilon_n(t)=1$), it will transmit the local model parameter to the edge server with power $P_n(t)$ ($P_n(t)>0$) in the uplink. Otherwise, the $n$-th MD keeps silent ($P_n(t)=0$). We assume that the size of the local training parameters of all MDs is the same, which is denoted by $M$.\footnote{We assume all MDs have the same structures of the local network model and bit precision (typically floating point precision) of the local network parameters, respectively.}
The uplink transmission rate for the $n$-th MD is given by
\begin{equation}
R_{n}(t) = \epsilon_n(t)W{\log_2}\left( {1 + \frac{{{P_n(t)}h_n(t)}}{{{\sigma ^2}}}} \right),
\end{equation}
where $W$ is the bandwidth of subchannel between each MD and the server, and $\sigma^2$ is the power of the additive white Gaussian noise.
Moreover, the corresponding uplink transmission time is expressed as
\begin{equation}
\tau_n^{\text{com}}(t) =\frac{{\epsilon_n(t) M}}{{R_{n}(t)}}.
\end{equation}
The energy consumption of parameter uploading for the $n$-th MD can be expressed as
\begin{equation}
E_n^{\text{com}}(t) = {P_n(t)}\tau_n^{\text{com}}(t)=\frac{{{\epsilon_n(t)P_n(t)}M}}{{R_{n}(t)}}.
\end{equation}

Upon receiving the updated local parameters from the MDs, the edge server aggregates them into a global parameter and then broadcasts it to all MDs through a downlink broadcast channel. Assume that the bandwidth of the broadcast channel is sufficiently wide and the transmit power of the edge server is much higher than that of the MDs. Therefore, we ignore the downlink transmission time without much loss of generality.

\section{Constrained Markov Decision Process}
\label{Constrained Markov Decision Problem Formulation}
In this section, we design and analyze the joint scheduling problem of computing and communication resources in the FL network. In the FL system, sequential decisions on local training and parameter transmission needs to be made for each iteration. From (\ref{state of the battery from the current time slot}), we know that the remaining energy at the MD sides are correlated among adjacent iterations. We therefore formulate the joint computing and communication resource scheduling problem as a CMDP to maximize the long term system reward under energy and delay constraints.

\subsection{The composition of CMDP}
At the beginning of each iteration, each MD uploads its current local channel state and battery energy state to the edge server. Therefore, the edge server obtains global status information to take appropriate actions for all MDs. Once the decisions are made, the edge server will download the policy to each MD. Due to the extremely small size of data for state information and action decisions, we can ignore the transmission delay and transmission energy for upload of local states and transmission of the policy in the FL network.
The CMDP formulation consists of the following components:
\begin{itemize}
  \item \textbf{State:} We define the global state $\boldsymbol{S}(t)$ of the all MDs in the $t$-th iteration as $\boldsymbol{S}(t)=[\boldsymbol{h}(t),\boldsymbol{E}^{\text{sta}}(t)]$, which is composed of the current global channel state $\boldsymbol{h}(t)=[h_1(t),...,h_N(t)]$ and the current global remaining battery energy state $\boldsymbol{E}^{\text{sta}}(t)=[E_1^{\text{sta}}(t),...,E_N^{\text{sta}}(t)]$.
  \item \textbf{Action:} Let us denote the global action $\boldsymbol{A}(t)$ of all MDs in the $t$-th iteration by $\boldsymbol{A}(t)=[\boldsymbol{b}(t),\boldsymbol{\epsilon}(t),\boldsymbol{P}(t)]$, which consists of the number of bits of training data $\boldsymbol{b}(t)=[b_1(t),...,b_N(t)]$, the upload decision $\boldsymbol{\epsilon}(t)=[{\epsilon}_1(t),...,{\epsilon}_N(t)]$ and the transmission power $\boldsymbol{P}(t)=[P_1(t),...,P_N(t)]$.
  \item \textbf{Transition probability:} According to the dynamic energy queue given in (\ref{state of the battery from the current time slot}), the global remaining energy $\boldsymbol{E}^{\text{sta}}(t)$ under action $\boldsymbol{A}(t)$ is a controlled Markov chain with the transition probability of
      \begin{equation}\label{energy dynamic Markovian transition}
\begin{split}
&\text{Pr}[\boldsymbol{E}^{\text{sta}}(t+1)|\boldsymbol{E}^{\text{sta}}(t),\boldsymbol{A}(t)]\\
&=\prod_{n}\text{Pr}\Big[ E^{\text{arr}}_n(t)=E_n^{\text{sta}}(t+1)\\
&\qquad \qquad  \left.-\left[E_n^{\text{sta}}(t)-\left\lceil E_n^{\text{com}}(t)+ E_n^{\text{cop}}(t) \right\rceil\right]^{+}\right].
\end{split}
\end{equation}
Since the energy queue dynamic is affected by both the data training energy and communication energy, it is controlled by the actions $\boldsymbol{A}(t)=\left(\boldsymbol{b}(t),\boldsymbol{\epsilon}(t),\boldsymbol{P}(t)\right)$. Moreover, the global state transition probability is also Markovian, which is given by
\begin{equation}\label{global state transition probability}
\begin{split}
&\text{Pr}[\boldsymbol{S}(t+1)|\boldsymbol{S}(t),\boldsymbol{A}(t))]\\
&=\text{Pr}\left[\boldsymbol{h}(t+1)|\boldsymbol{S}(t),\boldsymbol{A}(t))\right]\\&\qquad \qquad \qquad \quad \times \text{Pr}\left[\boldsymbol{E}^{\text{sta}}(t+1)|\boldsymbol{S}(t),\boldsymbol{A}(t))\right]\\
&=\text{Pr}\left[\boldsymbol{h}(t+1)\right]\text{Pr}\left[\boldsymbol{E}^{\text{sta}}(t+1)|\boldsymbol{S}(t),\boldsymbol{A}(t))\right],
\end{split}
\end{equation}
where the second equation is due to the i.i.d. property of wireless channel.
  \item \textbf{Reward:} The model accuracy of the FL is difficult to quantify, and does not promise a closed-form. In most circumstances, one observes that the accuracy of FL training increases with the total size of local training data at each MD \cite{anh2019efficient,li2014efficient}. Hence, we define the reward of the n-th MD by the product of its local training data size and its upload decision, i.e., $\sum_{n=1}^N b_n(t) \epsilon_n(t)$. If the MD is unable to upload the training parameters ($\epsilon_n(t)=0$), then its reward in the current iteration is 0.
\end{itemize}

We assume that each training iteration is synchronized across the MDs with the duration of $\tau$. Thus, the total time for data training and transmission should not exceed the duration of one iteration, i.e.,
\begin{equation}\label{delay constrains}
\begin{split}
\tau_n^{\text{com}}(t)+\tau_n^{\text{cop}}(t)\leq \tau.
\end{split}
\end{equation}
Moreover, the energy used for local training and uploading should not exceed the remaining energy $E_n^{\text{sta}}(t)$ at the beginning of the $t$-th iteration, which is described by the energy causality constraint of
\begin{equation}\label{energy constrains}
\begin{split}
 \left\lceil E_n^{\text{com}}(t)+ E_n^{\text{cop}}(t)\right\rceil \leq E^{\text{sta}}_n(t).
\end{split}
\end{equation}
From (\ref{delay constrains}) and (\ref{energy constrains}), we see that there is a tradeoff between the computing and communication phases due to limited time and battery energy in each training iteration. According to (\ref{f_c alpha}), if the number of training data bits $b_n$ increases, the computing energy consumption will increase, which leaves less energy for the communication phase. In the meanwhile, according to (\ref{t_c_b}), by increasing the training bits number $b_n$, the local training time will increase, which leaves less time for communication. Due to the above tradeoff, each MD needs to allocate time and energy wisely between the computing and communication phases. For example, the probability that the total remaining energy  $E^{\text{sta}}_n(t)$ at the $n$-th MD equals zero can not exceed the energy outage probability constraint $\text{Pr}_n^{\text{out}}$, i.e.,
\begin{equation}\label{energy state constraint}
\begin{split}
\text{Pr}[E_n^{\text{sta}}(t)=0]\leq \text{Pr}_n^{\text{out}},
\end{split}
\end{equation}
Here, $E_n^{\text{sta}}(t)=0$ does not mean that the MD is completely powered off. We adopt a dedicated battery to support the energy harvesting circuit and the information signaling in each training iteration. The dedicated battery stores the energy that arrives at random in each iteration, and provides energy for information feedback, local training and parameter updates during the FL process. We thus assume that the MDs can exhaust its battery before the next recharge cycle \cite{Sudevalayam2011energy}. Furthermore, we assume the subchannels occupied by all MDs in the current iteration cannot exceed the number of channels $L$ in the system, i.e.,
\begin{equation}\label{subchannel allocation constraint}
\begin{split}
\sum_{n=1}^N \epsilon_n(t)\leq L.
\end{split}
\end{equation}

Due to the randomness of states in each iteration and the correlation of states across the iterations, the edge server needs to make sequential decisions on $b_n$, $\epsilon_n$ and $P_n$ along the time horizon. Without much loss of generality, we formulate the problem as an infinite horizon CMDP, resulting in the stationary policies which do not change with time. The definition of stationary control policy is given as follows.
\begin{definition}
\emph{(Stationary Control Policy)} A stationary control policy is a mapping $\boldsymbol{S}\rightarrow \boldsymbol{A}$ from the state space to the action space $\mathcal{S}$,
which is given by $\boldsymbol{\Omega}(\boldsymbol{S})=\boldsymbol{A}\in \boldsymbol{\mathcal{A}}$, $\forall \boldsymbol{S}\in \boldsymbol{\mathcal{S}}$.
\end{definition}
Hence, we denote the control policy of the all MD by $\boldsymbol{\Omega}(S)=\left(\boldsymbol{b},\boldsymbol{\epsilon},\boldsymbol{P}\right)$. Let $\boldsymbol{\Omega}$ be the stationary feasible control policy which should satisfy constraints (\ref{delay constrains}), (\ref{energy constrains}), (\ref{energy state constraint}) and (\ref{subchannel allocation constraint}).

\subsection{Constrained Markov Decision Process Problem}
The formulation of CMDP is given in \emph{Problem \ref{Constrained MDP Problem}}. The aim is to find the efficient control policy that optimized the total long-term average utility of all MDs under the energy outage constraints, transmit power constraints, delay constraints, energy causality constraints and channel constraints.
\begin{problem}
\label{Constrained MDP Problem}
\emph{(CMDP Problem)}
\begin{align}
\mathop{\max}_{\boldsymbol{\Omega}}\ \qquad &\mathcal{U}(\boldsymbol{\Omega}) =\lim_{T\rightarrow \infty }\frac{1}{T}\sum_{t=1}^T \mathbb{E}^{\boldsymbol{\Omega}} \left[\sum_{n=1}^N b_n(t)\cdot \epsilon_n(t) \right]£¬\label{original constrained MDP}\\
\quad\text{s.t.}\qquad
&\text{Pr}[E_n^{\text{sta}}(t)=0]\leq \text{Pr}_n^{\text{out}},
 \tag{\ref{original constrained MDP}{a}}\label{constraint1 a}\\
&0\leq P_n(t)\leq P_n^{\text{max}}, \tag{\ref{original constrained MDP}{b}}\\
&\epsilon_n(t)\in \{0,1\},\notag\\
& (\ref{delay constrains}),\ (\ref{energy constrains}) \ \text{and} \ (\ref{subchannel allocation constraint}),\quad \forall n, \notag
\end{align}
\end{problem}
where the expectation $\mathbb{E}^{\boldsymbol{\Omega}}[\cdot]$ is taken with respect to the steady-state distribution induced by the control policy $\boldsymbol{\Omega}$, and $P_n^{\text{max}}$ is the maximum allowable transmit power of the $n$-th MD. Besides, the constraint (\ref{constraint1 a}) can be redescribed as a long-term description of the energy outage probability constraint, i.e.,
\begin{equation}
\lim_{T \rightarrow \infty}\frac{1}{T}\sum_{t=1}^T \mathbb{E}^{\boldsymbol{\Omega}} \left[\boldsymbol{1}[E^{\text{sta}}_n(t)=0] \right] \leq \text{Pr}_n^{\text{out}}.
\end{equation}
$\boldsymbol{1}[\cdot]$ is an indicator function that takes on a value of 1 when the battery energy is exhausted at the $n$-th MD. The objective function (\ref{original constrained MDP}) in \emph{Problem \ref{Constrained MDP Problem}} is the long-term average total utility of all MDs under the control policy $\boldsymbol{\Omega}$. In the following analysis, we decompose \emph{Problem \ref{Constrained MDP Problem}} into two stages. In stage one, we first omit the short term constraints in \emph{Problem \ref{Constrained MDP Problem}} and simplify the problem as follows,
\begin{problem}
\label{Simplified Constrained MDP Problem}
\emph{(Simplified CMDP Problem)}
\begin{align}
\mathop{\max}_{\boldsymbol{\Omega}}&\qquad \mathcal{U}(\boldsymbol{\Omega}) \label{original constrained MDP2}\\
\text{s.t.}\quad
&\lim_{T \rightarrow \infty}\frac{1}{T}\sum_{t=1}^T \mathbb{E}^{\boldsymbol{\Omega}} \left[ \boldsymbol{1}[E^{\text{sta}}_n(t)=0] \right] \leq \text{Pr}_n^{\text{out}}, \forall n,\tag{\ref{original constrained MDP2}{a}} \label{constraint2 a}\\
& \boldsymbol{\Omega}\in \boldsymbol{\mathcal{D}}(t), \forall t, \tag{\ref{original constrained MDP2}{b}}
\end{align}
\end{problem}
where $\boldsymbol{\mathcal{D}}(t)$ means the feasible region of the short term constraints in the learning iteration $t$. The Lagrange function of \emph{Problem \ref{Simplified Constrained MDP Problem}} is given by
\begin{equation}\label{the Lagrange function of Problem}
\begin{split}
L(\boldsymbol{\Omega},\boldsymbol{\gamma})
=\lim_{T \rightarrow \infty}\frac{1}{T}\sum_{t=1}^T \mathbb{E}^{\boldsymbol{\Omega}} \left[ g(\boldsymbol{S}(t),\boldsymbol{\Omega},\boldsymbol{\gamma})\right],
\end{split}
\end{equation}
where
\begin{equation}
\begin{split}
&g(\boldsymbol{S}(t),\boldsymbol{\Omega},\boldsymbol{\gamma})\\&= \sum_{n=1}^N \left(b_n(t) \epsilon_n(t)-\gamma_n \boldsymbol{1}[E_n^{\text{sta}}(t)=0]\right)
+\gamma_n \text{Pr}_n^{\text{out}}.
\end{split}
\end{equation}
And, the corresponding Lagrange dual function $G(\boldsymbol{\gamma})$ is given by
\begin{equation}\label{the Lagrange dual function of Problem}
\begin{split}
G(\boldsymbol{\gamma})
=\max_{\boldsymbol{\Omega}}\lim_{T \rightarrow \infty}\frac{1}{T}\sum_{t=1}^T \mathbb{E}^{\boldsymbol{\Omega}} \left[g(\boldsymbol{S}(t),\boldsymbol{\Omega},\boldsymbol{\gamma})\right],
\end{split}
\end{equation}
There exists Lagrange multipliers $\boldsymbol{\gamma}\succeq 0$ such that $\boldsymbol{\Omega}^*$ maximizes the Lagrange function $L(\boldsymbol{\Omega},\boldsymbol{\gamma})$. And we can get the following problem as,
\begin{problem}
\label{Lagrange dual problem}
\emph{(Lagrange Dual Problem)}
\begin{align}
L^*&= \min_{\boldsymbol{\gamma}} \max_{\boldsymbol{\Omega}} L(\boldsymbol{\Omega},\boldsymbol{\gamma})\\
&\text{s.t.}\quad \boldsymbol{\gamma}\succeq 0 \notag,\\
& \qquad\boldsymbol{\Omega}\in \boldsymbol{\mathcal{D}}(t), \forall t.\notag
\end{align}
\end{problem}According to \cite{Lei2016Delay1}, there exists an optimal control policy $\boldsymbol{\Omega}^*$ and a series of non-negative Lagrangian multipliers $\boldsymbol{\gamma}^*$ such that $\boldsymbol{\Omega}^*$ maximizes the Lagrange function $L(\boldsymbol{\Omega}^*,\boldsymbol{\gamma}^*)$, and the inequality condition holds:
\begin{equation}\label{saddle point condition}
\begin{split}
L(\boldsymbol{\Omega},\boldsymbol{\gamma}^*) \leq L(\boldsymbol{\Omega}^*,\boldsymbol{\gamma}^*) \leq L(\boldsymbol{\Omega}^*,\boldsymbol{\gamma}),\\
\forall \boldsymbol{\Omega}, \quad \forall\boldsymbol{\gamma}\succeq 0.
\end{split}
\end{equation}
$\boldsymbol{\Omega}^*$ and $\boldsymbol{\gamma}^*$ are the original optimal solution and the dual optimal solution, respectively. \emph{Problem \ref{Simplified Constrained MDP Problem}} can be regarded as an infinite-dimensional linear programming problem with the feasible region $\boldsymbol{\mathcal{D}}(t)$, which is a special type of convex problem. Thus, the duality gap between the original optimal and the dual optimal are $0$. The original optimal solution is obtained by solving the dual problem.

Generally, Bellman equation is a necessary condition for a dynamic programming to be optimized. Given Lagrange multipliers $\boldsymbol{\gamma}$, the classical
infinite-horizon average-utility CMDP problem \emph{Problem \ref{Simplified Constrained MDP Problem}} can be solved by the Bellman equation \cite{cui2012survey}. Thus, we can obtain the following equation,
\begin{equation}\label{Original Bellman Equation}
\begin{split}
&G(\boldsymbol{\gamma})+V(\boldsymbol{S}(t))=\max_{\boldsymbol{\Omega}}\Bigg\{g\left(\boldsymbol{S}(t),\boldsymbol{\Omega},\boldsymbol{\gamma}\right)+ \\ & \sum_{\boldsymbol{S}(t+1)}\text{Pr}\left(\boldsymbol{S}(t+1)|\boldsymbol{S}(t),\boldsymbol{\Omega}\right) V(\boldsymbol{S}(t+1))\Bigg\}\\
&\qquad \qquad \qquad \qquad \qquad \qquad  \qquad \forall \boldsymbol{S}(t), \quad t>0,
\end{split}
\end{equation}
where $V(\boldsymbol{S}(t))$ is the value function representing the average utility obtained by the control policy $\boldsymbol{\Omega}$ from each global state $[\boldsymbol{h}(t),\boldsymbol{E}^{\text{sta}}(t)]$. According to (\ref{global state transition probability}), we know that the channel states possess independent statistical characteristics, which is not affected by the control policy. We can further simplify the Bellman equation by taking the expectation of (\ref{Original Bellman Equation}) on the global channel state $\boldsymbol{h}(t)$.
\begin{lemma}\label{Equivalent Bellman Equation Lema}
\emph{(Equivalent Bellman Equation)} Given a series of Lagrange multipliers $\boldsymbol{\gamma}$, the objective function (\ref{the Lagrange function of Problem}) can be solved by the equivalent Bellman equation as follows:
\begin{equation}\label{Equivalent Bellman Equation}
\begin{split}
&G(\boldsymbol{\gamma})+V(\boldsymbol{E}^{\text{sta}}(t))=\max_{\boldsymbol{\Omega}(\boldsymbol{E}^{\text{sta}})}\Bigg\{ \overline{g}\left(\boldsymbol{E}^{\text{sta}}(t),\boldsymbol{\Omega}(\boldsymbol{E}^{\text{sta}})\right)+ \\& \sum_{\boldsymbol{E}^{\text{sta}}(t+1)}\text{Pr}\left(\boldsymbol{E}^{\text{sta}}(t+1)|\boldsymbol{E}^{\text{sta}}(t),\boldsymbol{\Omega}(\boldsymbol{E}^{\text{sta}})\right) V(\boldsymbol{E}^{\text{sta}}(t+1))\Bigg\}\\
&\qquad \qquad \qquad \qquad \qquad \qquad \qquad \qquad \forall \boldsymbol{E}^{\text{sta}}(t), \quad t>0,
\end{split}
\end{equation}
where the expectation of the value function $V(\boldsymbol{h}(t),\boldsymbol{E}^{\text{sta}}(t))$ is
\begin{equation}
\begin{split}
V(\boldsymbol{E}^{\text{sta}}(t))&=\mathbb{E}_{\boldsymbol{h}(t)}[V(\boldsymbol{h}(t),\boldsymbol{E}^{\text{sta}}(t))].
\end{split}
\end{equation}
Similarly, by taking the expectation over the channel state, we have
\begin{equation}
\begin{split}
\overline{g}&\left(\boldsymbol{E}^{\text{sta}}(t),\boldsymbol{\Omega}(\boldsymbol{E}^{\text{sta}})\right) \\ \qquad &=\mathbb{E}_{\boldsymbol{h}(t)}\left[ g(\boldsymbol{h}(t),\boldsymbol{E}^{\text{sta}}(t),\boldsymbol{\Omega},\boldsymbol{\gamma})\right],
\end{split}
\end{equation}
and,
\begin{align}
&\text{Pr}(\boldsymbol{E}^{\text{sta}}(t+1)|\boldsymbol{E}^{\text{sta}}(t),\boldsymbol{\Omega}(\boldsymbol{E}^{\text{sta}})) \\&=\mathbb{E}_{\boldsymbol{h}(t)}\left[\text{Pr}(\boldsymbol{E}^{\text{sta}}(t+1)|\boldsymbol{h}(t),\boldsymbol{E}^{\text{sta}}(t),\boldsymbol{\Omega}(\boldsymbol{E}^{\text{sta}}) \right].\notag
\end{align}
Moreover, $\boldsymbol{\Omega}(\boldsymbol{E}^{\text{sta}})=\{\boldsymbol{\Omega}(\boldsymbol{h}(t),\boldsymbol{E}^{\text{sta}}(t))|\forall \boldsymbol{h}(t)\}$ is a policy set under a given global energy state $\boldsymbol{E}^{\text{sta}}(t)$ for all possible channel states.
\end{lemma}

From the equivalent Bellman equation (\ref{Equivalent Bellman Equation}), we notice that the equation is composed by a series of linear equations, where the dimensions of these equations depend on the number of value functions $V(\boldsymbol{E}^{\text{sta}}(t))$. Hence, for any global energy state $\boldsymbol{E}^{\text{sta}}(t)$ and the global channel state $\boldsymbol{h}(t)$, the optimal control policy $\boldsymbol{\Omega}^*$ in (\ref{the Lagrange function of Problem}) can be obtained by maximizing the right-hand side of the equation (\ref{Equivalent Bellman Equation}).

\section{Approximate Markov Decision Process And Stochastic Learning}
\label{approximate_Markov_decision process1}
In this section, we use approximate MDP and stochastic learning methods to analyze and simplify the resource scheduling problem, and design online algorithms to obtain the resource scheduling policy for the FL system.

\subsection{Approximate Markov Decision Process}
According to (\ref{Equivalent Bellman Equation}), the global energy state value function $V(\boldsymbol{E}^{\text{sta}}(t))$ is unknown, which holds a great difficulty for solving the control policy in the FL system.
Due to the existence of the huge state-action space, we are unable to get the value function with the conventional value iteration method. However, we can obtain the value function and develop a solution of the \emph{Problem \ref{Constrained MDP Problem}} by the value approximation and online stochastic learning. Assumed that we have obtained the value function $V(\boldsymbol{E}^{\text{sta}}(t))$ through value approximation and online stochastic learning. Thus, the MDP problem can be solved as follow.

\begin{problem}
\label{Optimal Partitioned Actions}
\emph{(Optimal Partitioned Actions)} For a given value function $V(\boldsymbol{E}^{\text{sta}}(t))$, find the optimal partitioned actions $\boldsymbol{\Omega}^*(\boldsymbol{E}^{\text{sta}}(t))$, which is satisfied to the Equivalent Bellman's equation
in (\ref{Equivalent Bellman Equation}). The optimal control policy can be rewritten as
\begin{equation}
\label{optimal control policy given state value function}
\begin{split}
\boldsymbol{\Omega}^*(\boldsymbol{E}^{\text{sta}}(t))&=\arg\max_{\boldsymbol{\Omega}(\boldsymbol{E}^{\text{sta}}(t))} \mathbb{E}_{\boldsymbol{h}(t)} \left\{  g(\boldsymbol{S}(t),\boldsymbol{\Omega}(\boldsymbol{E}^{\text{sta}}(t)),\boldsymbol{\gamma})\right.\\&+ \sum_{\boldsymbol{E}^{\text{sta}}(t+1)}\Pr\left(\boldsymbol{E}^{\text{sta}}(t+1)|\boldsymbol{h}(t),\boldsymbol{E}^{\text{sta}}(t),\right.\\ & \left. \qquad\qquad\left.\boldsymbol{\Omega}(\boldsymbol{E}^{\text{sta}}(t))\right) V(\boldsymbol{E}^{\text{sta}}(t+1))\right\},\\
\textrm{s.t.}\qquad &0\leq P_n(t)\leq P_n^{\text{max}},\\
&\epsilon_n(t)\in \{0,1\},\\
\qquad \quad&(\ref{delay constrains}),\ (\ref{energy constrains}) \ \text{and} \ (\ref{subchannel allocation constraint}), \quad \forall n.
\end{split}
\end{equation}
\end{problem}

Given the global state value function $V(\boldsymbol{E}^{\text{sta}}(t))$ and the realization of the global channel state $\boldsymbol{h}$, the \emph{Problem} \ref{Optimal Partitioned Actions} then becomes a static optimization problem.

\subsection{Stochastic Learning}
By the feature-based method, the energy state value function $V(\boldsymbol{E}^{\text{sta}})$ can be approximated by a linear form of the state value
function of the $n$-th MD $V_n(E_n^{\text{sta}})$. The global energy state value function $V(\boldsymbol{E}^{\text{sta}})$ and a series of Lagrange multipliers $\boldsymbol{\gamma}$ will be updated according to the current energy state and channel state information. Then the proposed linear approximation architecture for the global energy state value
function $V(E_n^{\text{sta}})$ is obtained by

\begin{equation}
\label{linear approximation architecture}
\begin{split}
V(\boldsymbol{E}^{\text{sta}})&=V(E_1^{\text{sta}},...E_n^{\text{sta}},...E_N^{\text{sta}})\\
&\approx \sum_{n=1}^{N}\sum_{l\in \boldsymbol{Q}_n}V_n(l)\boldsymbol{I}[E_n^{\text{sta}}=l]=\boldsymbol{W}^T\boldsymbol{F}(\boldsymbol{E}^{\text{sta}}),
\end{split}
\end{equation}
where $\boldsymbol{Q}_n$ means energy state set of the $n$-th MD, that is $\boldsymbol{Q}_n=\{0,1,2,...,E_n^{\text{max}}\}$. The parameter vector $\boldsymbol{W}$ and the feature
$\boldsymbol{F}(\boldsymbol{E}^{\text{sta}})$ can be elaborated as,
\begin{equation}
\label{parameter vector}
\boldsymbol{W}=\left[V_1(0),...V_1 (E_1^{\text{max}}),...V_N(0),...V_N (E_N^{\text{max}})\right]^T,
\end{equation}
and
\begin{equation}
\label{feature vector}
\begin{split}
\boldsymbol{F}(\boldsymbol{E}^{\text{sta}})=\left[\boldsymbol{I}[E_1^{\text{sta}}=0],...\boldsymbol{I}[E_1^{\text{sta}}=E_1^{\text{max}}],\right.\\\left....\boldsymbol{I}[E_N^{\text{sta}}=0],...\boldsymbol{I}[E_N^{\text{sta}}=E_N^{\text{max}}]\right]^T.
\end{split}
\end{equation}
Thus, we can calculate the global energy state value function by the linear form of all MDs. According to the local energy state $E_n^{\text{sta}}$, the value function of global energy state
$\boldsymbol{E}^{\text{sta}}=\{E_1^{\text{sta}},...,E_N^{\text{sta}}\}$ can be expressed as
\begin{equation}
\label{local energy state global energy state value function}
V(\boldsymbol{E^{\text{sta}}})\approx\sum_{n=1}^N V_n(E_n^{\text{sta}}) \quad E_n^{\text{sta}}\in \boldsymbol{Q}_n.
\end{equation}
The global energy state value function $V(\boldsymbol{E}^{\text{sta}})$ is the same as the cardinality of the global energy state $\boldsymbol{E}=[E_1,...,E_N]$, and its number is $\prod_{n=1}^N(E_n^{\text{max}}+1)$. However, the number of the linear approximation energy state value function of all MDs is $\sum_{n=1}^N(E_n^{\text{max}}+1)$. Through linear approximation architecture, we exploit the state value function of each MD $V_n(E_n^{\text{sta}})$ with a small state space to represent the global energy state value function $V(\boldsymbol{E}^{\text{sta}})$ with huge state space.

According to \emph{Lemma} \ref{Equivalent Bellman Equation Lema} and the linear approximation architecture, we can obtain the following equations,

\begin{equation}
\label{Equivalent linear equation}
\begin{split}
&\mathbb{E}_{\boldsymbol{h}}\bigg\{\sum_{\boldsymbol{E}^{\text{sta}}(t+1)}\Pr\left(\boldsymbol{E}^{\text{sta}}(t+1)|\boldsymbol{h}(t),\boldsymbol{E}^{\text{sta}}(t)\right.,\\
&\qquad \qquad\qquad \left.\boldsymbol{\Omega}(\boldsymbol{h}(t),\boldsymbol{E}^{\text{sta}}(t))\right) V(\boldsymbol{E}^{\text{sta}}(t+1))\bigg\},\\
=&\mathbb{E}_{\boldsymbol{h}} \bigg\{\sum_{\boldsymbol{E}^{\text{sta}}(t+1)} \bigg( \prod_{n=1}^N \Pr (E_n^{\text{sta}}|\boldsymbol{h}(t),\boldsymbol{E}^{\text{sta}}(t),\\
&\qquad \qquad \boldsymbol{\Omega}(\boldsymbol{h}(t),\boldsymbol{E}^{\text{sta}}(t)))\sum_{n=1}^N V_n(E_n^{\text{sta}}(t+1))\bigg)\bigg\},\\
%
=&\mathbb{E}_{\boldsymbol{h}}\bigg\{\sum_{n=1}^N\sum_{E_n^{\text{sta}}(t+1)\in\boldsymbol{Q}_n} \Pr(E_n^{\text{sta}}(t+1)|\boldsymbol{h}(t),\boldsymbol{E}^{\text{sta}}(t),\\
&\qquad \qquad\qquad \boldsymbol{\Omega}(\boldsymbol{h}(t),\boldsymbol{E}^{\text{sta}}(t))) V_n(E_n^{\text{sta}}(t+1))\bigg\},\\
=&\mathbb{E}_{\boldsymbol{h}}\bigg\{\sum_{n=1}^N\sum_{A_n(t)}\Pr(A_n(t))V_n(E_n^{\text{sta}}(A_n(t),\Omega_n(\boldsymbol{S}(t))))\bigg\},
\end{split}
\end{equation}
where the post-action energy state of the $n$-th MD $E_n^{\text{sta}}(A_n(t),\Omega_n(\boldsymbol{S}(t)))$ can be defined as
\begin{equation}
\label{defination of post action}
\begin{split}
&E_n^{\text{sta}}(A_n(t),\Omega_n(\boldsymbol{S}(t)))=\\
&\min\Big\{\Big[E_n^{\text{sta}}(t)-\left\lceil E_n^{\text{com}}(t)+ E_n^{\text{cop}}(t) \right\rceil\Big]^+ +A_n(t),E_n^{\text{max}}\Big\}.
\end{split}
\end{equation}
The equation (\ref{Equivalent linear equation}) holds due to the state transition probability in (\ref{energy dynamic Markovian transition}) and the state update in (\ref{defination of post action}).
Thus, we can get the following optimal policy by (\ref{Equivalent linear equation}),
\begin{equation}
\label{linear approximation Bellman equation}
\begin{split}
&\boldsymbol{\Omega}^*(\boldsymbol{E}(t))=
\arg\max_{\boldsymbol{\Omega}(\boldsymbol{E}(t))} \mathbb{E}_{\boldsymbol{h}}\bigg\{g(\boldsymbol{S}(t),\boldsymbol{\Omega}(\boldsymbol{S}(t)),\boldsymbol{\gamma}) \\&\qquad+\sum_{n=1}^N\sum_{A_n(t)}\Pr(A_n(t))V_n(E_n^{\text{sta}}(A_n(t),\Omega_n(\boldsymbol{S}(t))))\bigg\}.
\end{split}
\end{equation}
According to the linear value approximation structure (\ref{local energy state global energy state value function}) and (\ref{linear approximation Bellman equation}), the control policy
problem can be re-written as the following problem.
\begin{problem}
\label{control policy2 problem}
\emph{(Equivalent control policy problem)}
\begin{equation}
\label{re-written the control policy2}
\begin{split}
&\max_{\boldsymbol{\Omega}^*} \mathbb{E}_{\boldsymbol{h}}\bigg\{g(\boldsymbol{S}(t),\boldsymbol{\Omega}(\boldsymbol{S}(t)),\boldsymbol{\gamma})\\
&\qquad +\sum_{n=1}^N\sum_{A_n(t)}\Pr(A_n(t))V_n(E_n^{\text{sta}}(A_n(t),\Omega_n(\boldsymbol{S}(t))))\bigg\},\\
&\quad\textrm{s.t.} \qquad 0\leq P_n(t)\leq P_n^{\text{max}},\\
&\qquad\qquad\epsilon_n(t)\in \{0,1\},\\
&\qquad\qquad(\ref{delay constrains}),\ (\ref{energy constrains}) \ \text{and} \ (\ref{subchannel allocation constraint}), \forall n.
\end{split}
\end{equation}
\end{problem}

Since $E^{\text{sta}}_n(A_n(t),\Omega_n(\boldsymbol{S}(t)))$ represents the update of the local energy state, we need to calculate the objective function of (\ref{re-written the control policy2}) for each local energy state,
and derive the objective function over all local energy states. To solve (\ref{re-written the control policy2}), we expand $V(E_n^{\text{sta}}(A_n(t),\Omega_n(\boldsymbol{S}(t))))$
in (\ref{re-written the control policy2}) using Taylor expansion as follows \cite{Bettesh_2006,Wang_Rui_2013}:
\begin{equation}
\label{energy state value function Taylor expansion}
\begin{split}
&V(E_n^{\text{sta}}(A_n(t),\Omega_n(\boldsymbol{S}(t))))
=V(E_n^{\text{sta}}(t))\\&\qquad\quad+\left(A_n(t)-\left\lceil E_n^{\text{com}}(t)+ E_n^{\text{cop}}(t) \right\rceil \right)V'(E_n^{\text{sta}}(t)),\\
&\textrm{where}\\
&V'(E^{\text{sta}}_n(t))=\left[V(E_n^{\text{sta}}(t)+1)-V(E_n^{\text{sta}}(t)-1)\right]/2.
\end{split}
\end{equation}
The optimization objective in (\ref{re-written the control policy2}) can be expressed as follow,
\begin{equation}
\label{re-written the control policy objective function}
\begin{split}
\max_{\boldsymbol{\Omega}} \quad &\mathbb{E}_{\boldsymbol{h}}\Big\{g(\boldsymbol{S}(t),\boldsymbol{\Omega}(\boldsymbol{S}(t),\boldsymbol{\gamma})\\
&+\sum_{n=1}^N\sum_{A_n(t)}\Pr(A_n(t))(V(E_n^{\text{sta}}(t))+(A_n(t)-\\ &\left.\left\lceil E_n^{\text{com}}(t)+ E_n^{\text{cop}}(t) \right\rceil \right)V'(E_n^{\text{sta}}(t)))\Big\}.
\end{split}
\end{equation}

In this way, we can obtain the equivalent optimization problem at the current iteration shown as follow, which is equivalent to (\ref{re-written the control policy2}),
\begin{equation}
\label{static mixed optimization}
\begin{split}
&\max_{\boldsymbol{b},\boldsymbol{\epsilon},\boldsymbol{P}} \quad g(\boldsymbol{S}(t),\boldsymbol{\Omega}(\boldsymbol{S}(t)),\boldsymbol{\gamma}) +\sum_{n=1}^N\sum_{A_n(t)}\Pr(A_n(t))  \\
&\qquad \times(A_n(t)-\left\lceil E_n^{\text{com}}(t)+ E_n^{\text{cop}}(t)\right\rceil)V'(E_n^{\text{sta}}(t)),\\
&\textrm{s.t.}\quad\qquad(\ref{delay constrains}),\ (\ref{energy constrains}) \ \text{and} \ (\ref{subchannel allocation constraint}),\\
&\qquad\qquad\epsilon_n(t)\in \{0,1\},\\
& \qquad\qquad\  0\leq P_n(t)\leq P_n^{\text{max}}, \quad \forall n,t,
\end{split}
\end{equation}
where (\ref{static mixed optimization}) is a static mixed variable optimization problem, in which $\boldsymbol{b}$ and $\boldsymbol{P}$ are continuous variables, while $\boldsymbol{\epsilon}$
are discrete variables. Besides, the ceiling operator $\left\lceil \cdot \right\rceil$ is difficult to handle, which brings great difficulties to the optimization problem. In order to solve the problem caused by the ceiling operator
$\left\lceil \cdot \right\rceil$, we introduce a series of auxiliary variables $\Delta E_n(t), \forall n$ to simplify the optimization problem. The optimization problem can be further described as follow:
\begin{equation}
\label{auxiliary variables optimization problem}
\begin{split}
&\max_{\boldsymbol{b},\boldsymbol{\epsilon},\boldsymbol{P}} \quad  g(\boldsymbol{S}(t),\boldsymbol{\Omega}(t),\boldsymbol{\gamma})+\sum_{n=1}^N\sum_{A_n(t)}\Pr(A_n(t)) \\
&\qquad \qquad\qquad\times(A_n(t)-\Delta E_n(t))V'(E_n(t))\\
&\quad\textrm{s.t.}\qquad(\ref{delay constrains})\ \text{and} \ (\ref{subchannel allocation constraint}),\\
&\qquad \qquad E_n^{\text{com}}(t)+ E_n^{\text{cop}}(t)- \Delta E_n(t) \leq 0,\\
&\qquad \qquad \Delta E_n(t) \in \{0,1,2,...,E_n(t)\},\\
&\qquad\qquad\epsilon_n(t)\in \{0,1\},\\
& \qquad\qquad\,  0\leq P_n(t)\leq P_n^{\text{max}}, \quad \forall n,
\end{split}
\end{equation}
where the auxiliary variable is 
\begin{equation}
\begin{split}
\Delta E_n(t)= &\left\lceil E_n^{\text{com}}(t)+ E_n^{\text{cop}}(t)\right\rceil\\
=&\left\lceil \alpha b_n(t){C_n}{f_n^2}+\frac{{{\epsilon_n(t)P_n(t)}d}}{{R_{n,s}(t)}} \right\rceil.
\end{split}
\end{equation}
Note that the constraints (\ref{subchannel allocation constraint}) describes the sub-channel constraints of all MDs, we ignore the constraints for the time being to simplify the optimization problem. Given a typical MD $n$, we can obtain the following optimization problem by further analysis and simplification of (\ref{static mixed optimization}),
\begin{equation}
\label{further static mixed optimization}
\begin{split}
&\max_{b_n,\epsilon_n,P_n} \quad \frac{\Delta E_n(t)-E_n^{\text{com}}(t)}{\alpha C_n f_n^2} \epsilon_n(t)-\Delta E_n(t)V'(E_n(t)),\\
&\qquad\textrm{s.t.}\qquad T_n^{\text{com}}(t)+T_n^{\text{cop}}(t)\leq \tau,\\
&\qquad\qquad \quad \Delta E_n(t) \in \{0,1,2,...,E_n(t)\},\\
&\qquad\qquad\quad\epsilon_n(t)\in \{0,1\},\\
& \qquad\qquad\quad 0\leq P_n(t)\leq P_n^{\text{max}}, \quad \forall n.
\end{split}
\end{equation}
From (\ref{further static mixed optimization}), we can draw the following conclusion obviously: if $\Delta E_n(t)\leq E_n^{\text{th}}(t)$, then $P_n(t)=0$
and $\epsilon_n(t)=0$, in which $E_n^{\text{th}}(t)$ means the threshold energy of the $n$-th MD at the current iteration, and it can be expressed as,
\begin{equation}
E_n^{\text{th}}(t)=\frac{\tau \sigma^2}{h_n(t)}\left(2^{\frac{d}{W\tau}}-1 \right).
\end{equation}
When $\Delta E_n(t)\leq E_n^{\text{th}}$, the energy consumed by the $n$-th MD at the current iteration
is insufficient to support uploading model parameters to the edge server within the iteration duration $\tau$.

Due to the first constraint of (\ref{further static mixed optimization}), we obtain the upper bound of energy consumption by the $n$-th MD at the iteration $t$, that is,
\begin{equation}
\Delta E_n(t) \leq (\tau-T^{\text{com}}_n(t))\alpha f_n^3+T^{\text{com}}_n(t)P_n(t).
\end{equation}
When $\frac{1}{\alpha C_n f_n^2}-V'(E_n(t))\leq0$, the energy consumption  $\Delta E_n(t)$ is $0$ obviously. Correspondingly, the transmission power $P_n(t)$ of the $n$-th MD, the transmission decision $\epsilon_n(t)$ of the $n$-th MD and the batch size of local training data $b_n(t)$ are all $0$. When $\frac{1}{\alpha C_n f_n^2} - V'(E_n(t)) >0$ and $\Delta E_n(t) > E_n^{\text{th}}(t)$, since the value of
$P_n(t)$ is related to the value of $\Delta E_n(t)$ and $\Delta E_n(t) \leq (\tau-T^{\text{com}}_n(t))\alpha f_n^3+T^{\text{com}}_n(t)P_n(t)$,
the energy consumption  $\Delta E_n(t)$ of the $n$-th MD take the maximum value, i.e.,
\begin{equation}
\Delta E_n(t)=\left(\tau-\frac{d}{R_{n,s}(t)}\right)\alpha f_n^3+\frac{d}{R_{n,s}(t)}P_n(t).
\end{equation}
Since $ P_n(t)\in (0,P^{\text{max}}_n ]$ and $\Delta E_n(t)$ increases monotonically as $P_n(t)$ increases, there is a maximum value of $\Delta E_n^{\text{max}}(t)$ as a function of $P_n(t)$,
which can be expressed as,
\begin{equation}
\Delta E_n^{\text{max}}(t)=\left(\tau-\frac{d}{R_{n,s}^{\text{max}}(t)}\right)\alpha f_n^3+\frac{d}{R_{n,s}^{\text{max}}(t)}P_n^{\text{max}}.
\end{equation}
Thus, when $\Delta E_n(t)>\Delta E_n^{\text{max}}(t)$,
the transmit power $P_n(t)$ is $P_n^{\text{max}}$, $\epsilon_n(t)=1$ and the batch size $b_n(t)$ for local training can be expresses as following,
\begin{equation}
\label{ local training data of mobile device Emax1}
\begin{split}
b_n(t)=\frac{\Delta E^{\text{max}}_n(t)-E_{n,\text{max}}^{\text{com}}(t)}{\alpha C_n f_n^2},
\end{split}
\end{equation}
where,
\begin{equation}
\label{ local training data of mobile device Emax2}
\begin{split}
E_{n,\text{max}}^{\text{com}}(t)=\frac{dP_n^{\text{max}}}{W \log_2 \left(1+\frac{P_n^{\text{max}}h_n(t)}{\sigma^2}\right)}.
\end{split}
\end{equation}
Then for $E_n^{\text{th}}(t) <\Delta E_n(t)\leq \Delta E^{\text{max}}_n(t) $, according to the relationship between $\Delta E_n(t)$ and $P_n(t)$, we can express $P_n(t)$ by $\Delta E_n(t)$, which has shown as follow,
 \begin{equation}
 \label{relationshio between E anP}
 \begin{split}
 &P_n(t)=\frac{\textrm{lambertW}\left(\frac{B_n(t)}{Z_n(t)}e^{\frac{C_n(t)}{Z_n(t)}}\right)}{\frac{B_n(t)}{Z_n(t)}h_n(t)}-\frac{\sigma^2}{h_n(t)},\\
 &\textrm{where,}\\
 &B_n(t)=-\frac{d}{h_n(t)},\\
 &C_n(t)=\frac{W(\Delta E_n(t)-\alpha f_n^3\tau)}{\ln2}\ln\sigma^2-\frac{d \sigma^2}{h_n(t)}-\alpha f_n^3 d,\\
 &Z_n(t)=\frac{W(\Delta E_n(t)-\alpha f_n^3\tau)}{\ln2}.
 \end{split}
 \end{equation}
And $\epsilon_n(t)=1$, $\textrm{lambertW}$ means Lambert W Function, which is the inverse function of $f(w)=w\cdot \exp(w)$. And it is a special function that cannot be represented by an expression. From this, we
get the objective function of the variable of $\Delta E_n$ by substituting $P_n(t)$ into the objective function in (\ref{further static mixed optimization}),
\begin{equation}
\label{objective function of E}
\begin{split}
&F_n(t)=\max_{\Delta E_n(t)} \frac{1}{\alpha C_n f_n^2}\bigg(\min \left\{\Delta E_n(t),\Delta E^{\text{max}}_n(t)\right\}\\
&   -\frac{dP_n(\Delta E_n(t))}{R_{n,s}(\Delta E_n(t))}\bigg)\epsilon_n(t)-\Delta E_n(V'_n(E_n(t))\\
&\textrm{s.t.} \quad\Delta E_n(t) \in \left\{0,1,...,\min \left\{E_n^{\text{sta}}(t),\left\lceil \Delta E^{\text{max}}_n(t)\right\rceil \right\}\right\}.
\end{split}
\end{equation}

\begin{algorithm}[h]
\label{static mixed variable optimization problem algorithm2}
    \caption{The pseudocode of the proposed static mixed variable optimization problem of all MDs}%
    \KwIn{input $\boldsymbol{E^{\text{sta}}}(t), \boldsymbol{h}(t), f_n, P_n^{\text{max}}, V_n'(t), \forall n$;}
    \KwOut{output result $\boldsymbol{P}^*(t),\boldsymbol{b}^*(t),\boldsymbol{\epsilon}^*(t)$;}
    \For {The $n$-th MD, $n\in {1,...,N}$}{
        Calculate threshold energy $E_n^{\text{th}}(t)$ and the maximum energy consumption $\Delta E_n^{\text{max}}(t)$ for calculation and communication in the $t$-th iterartion from input parameters\;
        \For {$\Delta E_n(t) \in \left\{0,1,...,\min \left\{E_n(t),\left\lceil \Delta E_n^{\text{max}}(t)\right\rceil\right\}\right\}$ } {
        \uIf { $\Delta E_n(t) \leq E_n^{\text{th}}(t)$}{
        $P_n(t)=0, b_n(t)=0, \epsilon_n(t)=0$\;}
        \uElseIf {$\frac{1}{\alpha C_n f_n^2}-V_n'(E_n(t))\leq0$}{
        $P_n(t)=0, b_n(t)=0, \epsilon_n(t)=0$\;}
        \uElseIf {$E_n^{\text{th}}(t) <\Delta E_n(t)\leq \Delta E^{\text{max}}_n(t)$}{
         The solution of $P_n(t)$ and $b_n(t)$ can refer to the formula (\ref{relationshio between E anP}) and $\epsilon_n(t)=1$\;}
        \uElse{
        $P_n(t)=P_n^{\text{max}}$ and the solution of $b_n(t)$ can refer to the formula (\ref{ local training data of mobile device Emax1}) and $\epsilon_n(t)=1$\;}
        $\textbf{end if}$\\
        Substituting the values of $P_n(t)$ and $\epsilon_n(t)$ that have been obtained into the objective function of (\ref{further static mixed optimization}) \;
        }
        By searching in the $\left\{0,1,...,\min \left\{E_n(t),\left\lceil \Delta E^{\text{max}}_n(t) \right\rceil \right\}\right\}$, the optimal energy consumption value $\Delta \hat{E}_n(t)$ of the $n$-th MD for the maximum value of the objective
        function in (\ref{objective function of E}) can be found, and $\hat{P}_n(t),\hat{b}_n(t),\hat{\epsilon}_n(t)$ can be calculated.
        }
        \uIf {$\parallel\hat{\boldsymbol{\epsilon}}\parallel_1\leq L$}{
        The global optimal solution is equal to the solution obtained by the respective MD, i.e., $\boldsymbol{P}^*(t)=\hat{\boldsymbol{P}}(t)$, $\boldsymbol{b}^*(t)=\hat{\boldsymbol{b}}(t)$.}
        \uElse{The edge server will select $L$ MDs with the largest $F_n(t)$ for FL training. If the $n$-th MD is selected by the server, then $P^*_n(t)=\hat{P}_n(t), \epsilon^*_n(t)=\hat{\epsilon}_n(t)$ and $b^*_n(t)=\hat{b}_n(t)$, otherwise $P^*_n(t)=0, \epsilon^*_n(t)=0$ and $b^*_n(t)=0$.}
        $\textbf{end if}$\\

\end{algorithm}
By searching in the $\left\{0,1,...,\min \left\{E_n(t),\left\lceil E^{\text{max}}_n(t)\right\rceil \right\}\right\}$, we can find the optimal energy consumption value $\Delta \hat{E}_n(t)$ of the $n$-th MD for the maximum value of the objective
function in (\ref{objective function of E}). For convenience, we assume $0/0 = 0$ for the term of $\frac{dP_n(\Delta E_n(t))}{R_{n,s}(\Delta E_n(t))}$ in this paper. Recalling the sub-channel constraint (\ref{subchannel allocation constraint}) that we ignored earlier, we will analyze it.
According to (\ref{objective function of E}), we get the optimal objective function $F_n(t)$ of each MD in the current iteration. If $\parallel\hat{\boldsymbol{\epsilon}}\parallel_1\leq L$, the global optimal solution is equal to the solution obtained by the respective MD, i.e., $P^*_n(t)=\hat{P}_n(t), b^*_n(t)=\hat{b}_n(t)$. Then, when $\parallel\hat{\boldsymbol{\epsilon}}\parallel_1> L$, the edge server will select $L$ MDs with the largest $F_n(t)$ for FL training. If the $n$-th MD is selected by the edge server to upload parameters, then $P^*_n(t)=\hat{P}_n(t), \epsilon^*_n(t)=\hat{\epsilon}_n(t)$ and $b^*_n(t)=\hat{b}_n(t)$, otherwise $P^*_n(t)=0, \epsilon^*_n(t)=0$ and $b^*_n(t)=0$.  Algorithm \ref{static mixed variable optimization problem algorithm2} reports the pseudocode of the proposed static mixed variable optimization problem.

In the previous section, we assumed that the state value function $V(\boldsymbol{E}^{\text{sta}})$ has been given. However, we need to know the state value function of each MD accurately
so that we can make efficient control decisions in the FL process. We utilize stochastic learning and propose a distributed online algorithm to estimate the value function $V(\boldsymbol{E}^{\text{sta}})$
and the Lagrange multipliers $\boldsymbol{\gamma}$ based on the current state. The updates of the value function $V$ at the end of the iteration $t$ can be given
by (\ref{t time slot value function}).

\begin{algorithm}[h]
    \caption{The specific flow of the stochastic learning}%
    \label{The specific process of stochastic learning}
        Initialize the respective energy state value function vectors $\boldsymbol{V}^0$ and the Lagrange multiplier vectors  $\boldsymbol{\gamma}^0$ of all MDs \;
        Based on the observed local states, a series of parameters and the local energy value functions $\boldsymbol{V}^t$ of each MD, the control action can be  calculated
        by Algorithm \ref{static mixed variable optimization problem algorithm2} at the beginning of the iteration $t$\;
        Based on the observed local states, the control actions and the instantaneous rewards of the system, the energy state value function $\boldsymbol{V}^{t+1}$ and Lagrange multiplier vectors
        $\boldsymbol{\gamma}^{t+1}$ can be updated by (\ref{t time slot value function}), (\ref{delta value function}) and (\ref{lM update11})\;
        If $\parallel \boldsymbol{V}^{t+1}-\boldsymbol{V}^{t} \parallel < \delta_v$ and $\parallel \boldsymbol{\gamma}^{t+1}-\boldsymbol{\gamma}^{t} \parallel<\delta_{\gamma}$, stop;
        otherwise, set $t=t+1$ and go back to step 2.
\end{algorithm}

\begin{equation}
\label{t time slot value function}
\begin{split}
V_n^{t+1}(l)=
\begin{cases}
(1-\epsilon_v^t)V_n^{t}(l)+\epsilon_v^t\Delta V_n^{t+1}(l)  &\text{if}\quad \boldsymbol{E}^{t+1}_n=l\\
V_n^{t}(l)   &\text{if}\quad \boldsymbol{E}^{t+1}_n\neq l
\end{cases}
\end{split}
\end{equation}
where $\Delta V_n^{t}(l)$ is expressed in (\ref{delta value function}),

\begin{equation}
\label{delta value function}
\begin{split}
\Delta V_n^{t+1}(l)&=b_n(t)\epsilon_n(t)-\gamma_n^t \boldsymbol{1}[l=0]\\ +&\sum_{A_n}\left\{\Pr(A_n)(V_n^t(l(\Delta E_n^{t+1},A_n))-V_n^t(l(A_n)))\right\}
\end{split}
\end{equation}
Moreover, the Lagrange multipliers updates at per MD are given by
\begin{equation}
\label{lM update11}
\begin{split}
\gamma_n^{t+1}=[\gamma_n^t+\epsilon^t_{\gamma}(\boldsymbol{1}[E_n^{t+1}=0]-\text{Pr}_n^{\text{th}})]^+
\end{split}
\end{equation}
In the above equations, $\left(\{\epsilon_v^t\},\{\epsilon_{\gamma}^t\} \right)$ are the sequences of iteration size, which satisfy,
\begin{equation}
\begin{split}
&\sum_{t=0}^{\infty}\epsilon_v^t=\infty,\ \epsilon_v^t>0, \  \lim_{t\rightarrow \infty}\epsilon_v^t =0,\\
&\sum_{t=0}^{\infty}\epsilon_{\gamma}^t=\infty,\ \epsilon_v^{\gamma}>0, \  \lim_{t\rightarrow \infty}\epsilon_v^{\gamma} =0,\\
&\sum_{t=0}^{\infty}\left[(\epsilon_v^t)^2+(\epsilon_{\gamma}^t)^2\right]<\infty, \ \text{and}\  \lim_{t\rightarrow \infty}\frac{\epsilon_{\gamma}^t}{\epsilon_v^t}=0.
\end{split}
\end{equation}
The specific process of stochastic learning can refer to Algorithm \ref{The specific process of stochastic learning}.

\subsection{Convergence Analysis}
We need to provide effective analysis for the convergence of the online stochastic learning algorithm, which is shown in Algorithm \ref{The specific process of stochastic learning}.
From the previous section, we notice that there are two different step size sequences $\{\epsilon_v^t\}$ and $\{\epsilon_{\gamma}^t\}$ in the stochastic learning process, which
are used for the update of state value functions of MDs and Lagrange Multipliers respectively. Since the update of the Lagrangian multiplier $\boldsymbol{\gamma}$ and the update of
the value function $\boldsymbol{V}$ occur simultaneously and $\epsilon_{\gamma}^t=\boldsymbol{o}(\epsilon_v^t)$, we can obtain $\gamma^{t+1}-\gamma^t=\boldsymbol{o}(\epsilon_v^t)$.
Therefore, we consider that the Lagrangian multipliers does not change when the state value function is updated. Therefore, we assume that the Lagrangian multipliers $\boldsymbol{\gamma}^t$ keep
static when the value functions of the mobile devices are updated in (\ref{t time slot value function}).

The relationship between the the global value function vector $\boldsymbol{V}$ and the parameter vector $\boldsymbol{W}$ can be expressed as,
\begin{equation}
\label{relationship between the the global value function vector and the parameter vector}
\boldsymbol{V}=\boldsymbol{MW}  \qquad \text{and} \qquad  \boldsymbol{W}=\boldsymbol{M}^{\dagger} \boldsymbol{V},
\end{equation}
in which $\boldsymbol{M} \in \mathbb{R}^{| I_S|\times \sum_{n=1}^N\left( E_n^{\text{max}}+1\right)}$ with the $k$th row ($k=1,2,...,|I_S|$) equal to $\boldsymbol{F}(\boldsymbol{E}^k)$, where
$\boldsymbol{E}^k$ is the $k$th global energy state and $|I_S|$ is the cardinality of the system state. In addition, $\boldsymbol{M}^{\dagger}  \in \mathbb{R}^{\sum_{n=1}^N\left( E_n^{\text{max}}+1\right)\times |I_S|}$ means the mapping
matrix from $\boldsymbol{V}$ to $\boldsymbol{W}$, which is the inverse mapping of the first equation of (\ref{relationship between the the global value function vector and the parameter vector}).
We then have the following convergence lemma on the local state value function for each MD in the stochastic learning.

\begin{lemma}
\label{convergence of local state value function}
\emph{(Convergence of State Value Function of each MD):} The convergence performance of the state value function can be expressed mathematically as follows.
\begin{enumerate}
  \item The update of the state value function vector converge almost surely for any given initial parameter vector $\boldsymbol{W}^0$ and Lagrange multiplier $\gamma$, which
  can be expressed as
  \begin{equation}
  \label{update of the state value function vector converge 1}
  \lim_{t \rightarrow \infty} \boldsymbol{W}^t(\gamma) = \boldsymbol{W}^{\infty}(\gamma).
  \end{equation}
  \item The local steady-state value function vector $\boldsymbol{W}^{\infty}$ satisfies the vector form of the following steady equivalent Bellman equation,
  \begin{equation}
  \label{local steady-state value function vector}
  \theta \boldsymbol{I} + \boldsymbol{W}^{\infty}(\gamma)= \boldsymbol{M}^{\dagger} \boldsymbol{T}\left(\gamma, \boldsymbol{M}\boldsymbol{W}^{\infty}(\gamma)\right),
  \end{equation}
  where $\boldsymbol{I}$ is a $\sum_{n=1}^N\left( E_n^{\text{max}}+1\right)\times 1$ vector whose elements are all equal to $1$, $\boldsymbol{T}$ represents a function mapping, which
  can be defined as,
  \begin{equation}
  \label{T function mapping}
  \boldsymbol{T}(\gamma,\boldsymbol{V})=\max_{\Omega} \left \{ \overline{\boldsymbol{g}}\left(\gamma,\Omega\right) + \boldsymbol{P}(\Omega) \boldsymbol{V}\right \}
  \end{equation}
  where $\overline{\boldsymbol{g}}\left(\gamma,\Omega\right)$ is a $\sum_{n=1}^N\left( E_n^{\text{max}}+1\right)\times 1$ vector of function $\overline{g}\left(\boldsymbol{E},\Omega(\boldsymbol{E})\right)$,
  which is defined in (\ref{Equivalent Bellman Equation}). $\boldsymbol{P}(\Omega)$ is the matrix form of transition probability $\text{Pr}(\boldsymbol{E}^{t+1}|\boldsymbol{E}^t,\Omega)$
  defined in (\ref{Equivalent Bellman Equation}).
\end{enumerate}
\emph{Proof:}\quad Following \cite{Wang_Rui_2013}, we briefly explain the Lemma. Since we consider the stochastic channels, where the
channel gain varies within the interval, it is easy to see that each state will be updated comparably often in the asynchronous learning algorithm. Quoting the conclusion in \cite{Wang_Rui_2013}, the convergence property of the asynchronous update and synchronous
update is the same. Therefore, we just consider the convergence of related synchronous version for simplicity in this proof.
According to the definition of parameter vector $\boldsymbol{W}$ and the bounded per-MD value function $V_n$, it is clearly that the update on the per-MD value function vector is equivalent to the update on the parameter vector and to prove the convergence of the Lemma is equivalent to prove the convergence of update on the parameter vector $\boldsymbol{W}$. The proof of details can refer to \cite{Wang_Rui_2013}.

\end{lemma}

Due to $\epsilon_{\gamma}^t=\boldsymbol{o}(\epsilon_v^t)$, the ratio of step sizes between state value function and Lagrange Multiplier can be expressed as
$\frac{\epsilon_{\gamma}^t}{\epsilon_v^t}\rightarrow 0$ during the Lagrange Multiplier¡¯ update in (\ref{lM update11}), and the updates of the local state value function are much
faster than the Lagrange Multiplier. Thus, the Lagrange Multiplier can be consider as quasi-invariant during the update of the local state value functions of each MD, and
the update of the Lagrange Multiplier will trigger another update process of the local state value function of each MD. According to \cite{Borkar_1997}, we can obtain that
$\lim_{t \rightarrow \infty} ||V^t_n-V^{\infty}_n(\gamma^t)||=0$, in which $V^{\infty}_n(\gamma^t)$ means the converged local state value function of
$n$th MD with Lagrange Multiplier $\gamma^t$. Therefore, the update of the local state value function can be considered as almost constant during the Lagrange Multiplier¡¯s update.
Then, we need to the convergence lemma of the Lagrange Multipliers.

\begin{lemma}
\label{Convergence of the Lagrange Multiplier}
\emph{(Convergence of the Lagrange Multipliers):} The iteration on the Lagrange Multipliers $\boldsymbol{\gamma}$ converges almost surely to the set of minimum of $G(\boldsymbol{\gamma})$
in (\ref{the Lagrange dual function of Problem}). Supposing that the Lagrange Multipliers converge to $\boldsymbol{\gamma}^*$, then $\boldsymbol{\gamma}^*$
satisfies the average energy outage constraint in (\ref{energy state constraint}).

\emph{Proof:}\quad Quoting to \cite[Lemma 4.2]{Borkar_2005}, $-G(\boldsymbol{\gamma})$ is a concave and continuously differentiable except at finitely many points where both right and left derivatives exist.
Thus, $G(\boldsymbol{\gamma})$ is a convex function of $\gamma$. Since the energy consumption policy of each MD is discrete, we can obtain that $\Omega^*(\gamma)=\Omega^*(\gamma+\Delta_{\gamma}
)$, i.e., $\nabla_{\gamma}=(\Omega^*(\gamma+\Delta_{\gamma})-\Omega^*(\gamma))/\Delta_{\gamma}=0$. Thus, $\partial G(\gamma^t)/\partial\gamma^t$ can be expressed as $\partial G(\gamma^t)/\partial\gamma^t=
\mathbb{E}^{\Omega^*(\gamma^t)}\{\text{Pr}_n^E-\boldsymbol{1}[E_n(t)=0]\}$, where $\Omega^*(\gamma^t)=\arg\max_{\Omega} G(\gamma^t)$. By the standard stochastic approximation theorem \cite{ljung2012stochastic},
the dynamics of the Lagrange Multiplier update can be represented by ordinary differential equation (ODE). According to \cite{Lei_2016}, we know that that the ODE equals to
$\partial G(\gamma^t)/\partial\gamma^t$. Thus, the aforementioned ODE will converge to $\partial G(\gamma^t)/\partial\gamma^t=0$, i.e., the the
average energy outage constraints are satisfied. $\hfill\blacksquare$
\end{lemma}

According to Lemma \ref{convergence of local state value function} and \ref{Convergence of the Lagrange Multiplier}, the iteration on local state value function
and the Lagrange Multipliers in Algorithm \ref{The specific process of stochastic learning} will converge.

\section{Simulation And Discussion}
\label{sumulation1}
In this section, we evaluate the performance of the proposed algorithm using numerical results. In the simulations, all MDs are
randomly distributed in a fixed region. We set the bandwidth of channel between each MD and the edge server as $0.1\text{MHz}$. The number of CPU cycles $C$ for each MD to perform local model training of unit data sampling takes
range from $10^{10} \text{cycle/unit}$ to $1.9*10^{10} \text{cycle/unit}$. The simulation parameters are detailed in Table \ref{Parameters in Simulations}.

\begin{table}[h]
\caption{Parameters in Simulations}
\label{Parameters in Simulations}
\center
\begin{tabular}{p{5cm}|p{2.5cm}}
\hline
\hline
Nations & Values  \\
\hline
The number of MDs $N$ & $10$ \\
\hline
The number of channels in FL system $L$ & $5$ \\
\hline
Channel bandwidth $W$ & $0.1\text{MHz}$ \\
\hline
The number of CPU cycles $C$ per unit data sampling & $[10^{10}, 1.9\times10^{10}]$ $\text{cycles/unit}$\\
\hline
Each iteration duration $\tau$ & $10\text{s}$\\
\hline
Computation capacity $f_n$ of the MD & $[2\times10^9, 4\times10^9]$ \text{cycles/s}\\
\hline
The size of local parameter for each MD & $10^6 \text{bit}$\\
\hline
The effective capacitance parameter $\alpha$ & $10^{-28}$\\
\hline
The coefficient determined by machine learning model $\zeta$ & $1$\\
\hline
The upper limit value of the average energy outage $\text{Pr}_n^{\text{th}}$ & $4\%$\\
\hline
\hline
\end{tabular}
\end{table}
We compare our proposed approximate MDP solution with online stochastic learning with three other reference control algorithms. One is the CSI-based MDP algorithm, where
the edge server takes corresponding decisions based on the channel state only at the current iteration so as to optimize average utility of
all MDs. The second reference control algorithm is myopic method, which is a method that only considers the current utility. In myopic method,
the edge server never considers long-term utilities. The last reference control algorithm is a random
resource scheduling method, where the edge server takes random actions in the feasible regions. The performance of the proposed algorithm is evaluated by
averaging over 5000 experiments.

\begin{figure}

\centering
\includegraphics[height=6cm,width=8.5cm]{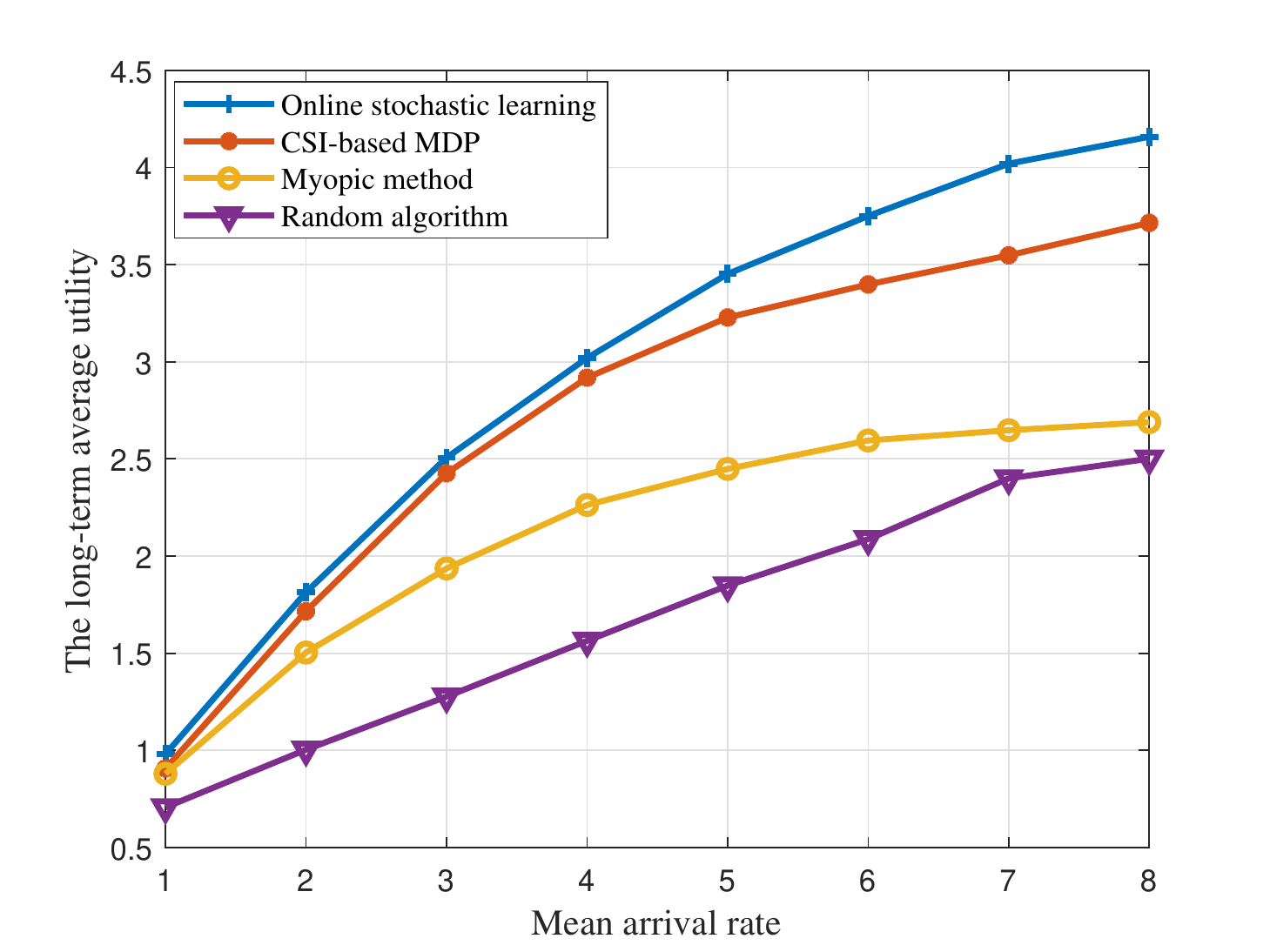}
\caption{The long-term average utility $\mathcal{U}$(MB) v.s. the mean arrive rate $\lambda$(J)of the random new arrived energy with $E^{\text{max}}=6\text{J}$. }
\label{fig:lambda_vs_utility2}
\end{figure}

Fig. \ref{fig:lambda_vs_utility2} illustrates the long-term average utility $\mathcal{U}$ v.s. the mean arrival rate $\lambda$ of the
random new arrived energy with $E^{\text{max}}=6\text{J}$. It can be observed that  the performance of the online stochastic learning algorithm is
better than the other reference algorithms for all the investigated average arrival rate $\lambda$. When the value of the mean arrival rate $\lambda$ is relatively small,
the performance of online learning is close to that of other reference algorithms, especially the CSI-based MDP algorithm. The cause of this phenomenon is twofold.
First, the small mean arrival rate $\lambda$ of energy will result in a limited energy level in the battery of the MD. Due to insufficient battery energy,
MDs are constrained with a small space of actions compared with those with sufficient battery power. The second reason is that a large amount of battery energy is used
for uploading local parameters. When the battery energy of MD is insufficient, the energy for local training of FL is smaller,
which leads to smaller long-term average utility. In contrast, when the battery level is high, MD's actions will
become more diverse, and more energy will be used for local training in FL process.

\begin{figure}
\centering
\includegraphics[height=6cm,width=8.5cm]{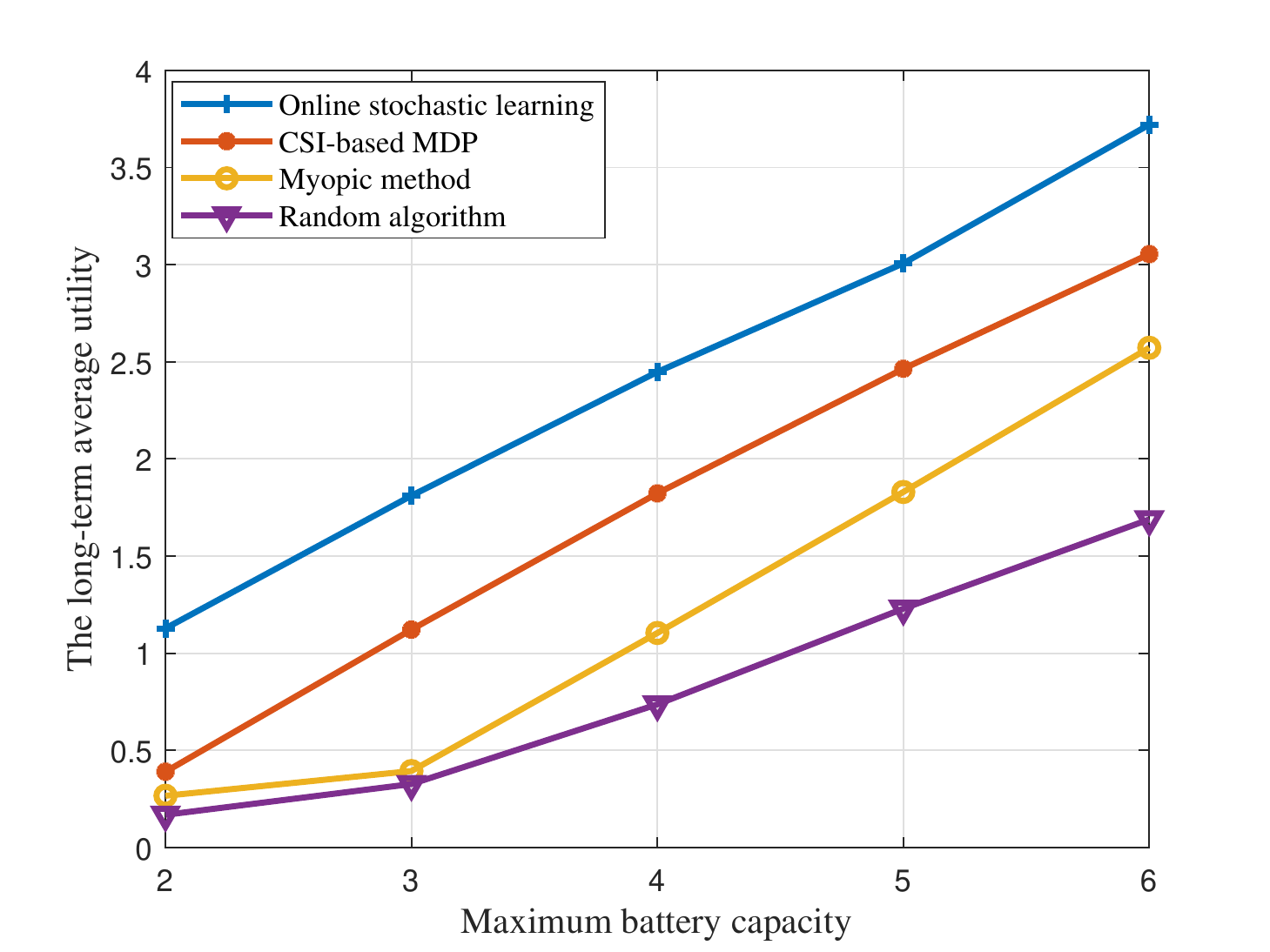}
\caption{The long-term average utility $\mathcal{U}$(MB) v.s. the maximum battery capacity $E^{\text{max}}$(J) of MD. }
\label{fig:capacity_vs_utility}
\end{figure}

Fig. \ref{fig:capacity_vs_utility} illustrates the long-term average utility $\mathcal{U}$ v.s. the maximum battery capacity $E^{\text{max}}$ of
MD. We observe that the long-term average utility increases as the maximum battery capacity $E^{\text{max}}$ of
MD increases in all algorithms,although the mean arrival rate $\lambda$ of the arrival energy has never changed. It indicates that the long-term
average utility $\mathcal{U}$ increases approximately linearly with the maximum battery capacity $E^{\text{max}}$ in our proposed algorithm.

\begin{figure}
\centering
\includegraphics[height=6cm,width=8.5cm]{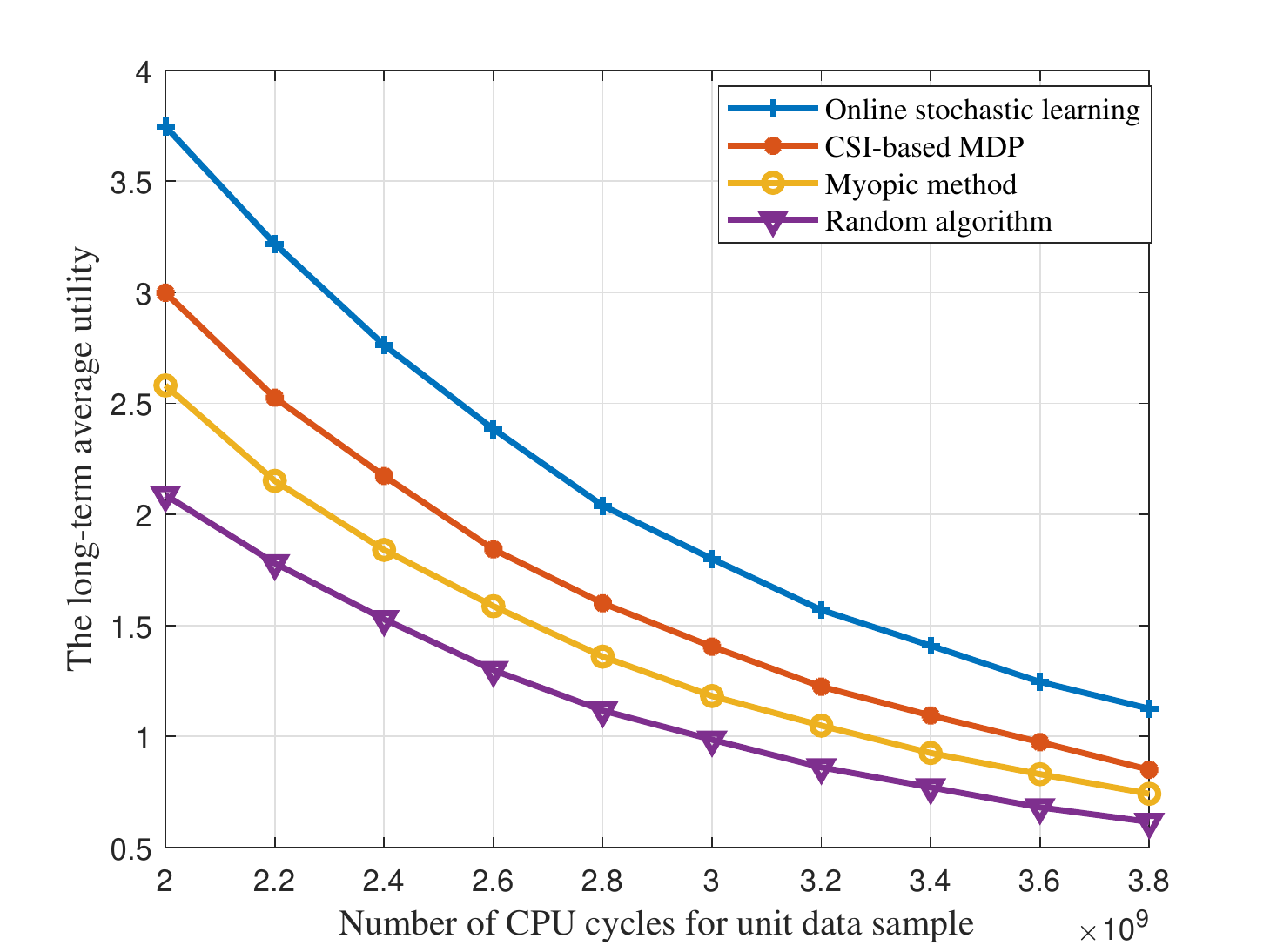}
\caption{The long-term average utility $\mathcal{U}$(MB) v.s. the number of CPU cycles for MD to perform local model training of unit data sampling $C$(cyc/MB).}
\label{fig:Cc_vs_utility}
\end{figure}

Fig. \ref{fig:Cc_vs_utility} shows the impact of the number of CPU cycles for MD to perform local model training of unit data sampling $C$ on
the long-term average utility $\mathcal{U}$. It is obvious that the long-term average utility decreases as the number of CPU cycles for unit
training data sampling $C$. In addition, as the number of CPU cycles for unit training data sampling $C$ continues to increase, performance differences
among different algorithms also decrease.  Fig. \ref{fig:fc_vs_utility} depicts the long-term average utility $\mathcal{U}$ v.s.
the computation capacity $f$ of MD. Intuitively, the MDs with more computing capacity will lead to a higher long-term
average utility $\mathcal{U}$. However, the opposite is true, it is caused by (\ref{f_c alpha}) and the limited energy of MD
in each iteration. In other words, a more powerful computing capacity requires more computing energy. Due to the limited amounts of energy available to MDs in each
iteration, MDs can only reduce the size of sampling data used for local training.

\begin{figure}
\centering
\includegraphics[height=6cm,width=8.5cm]{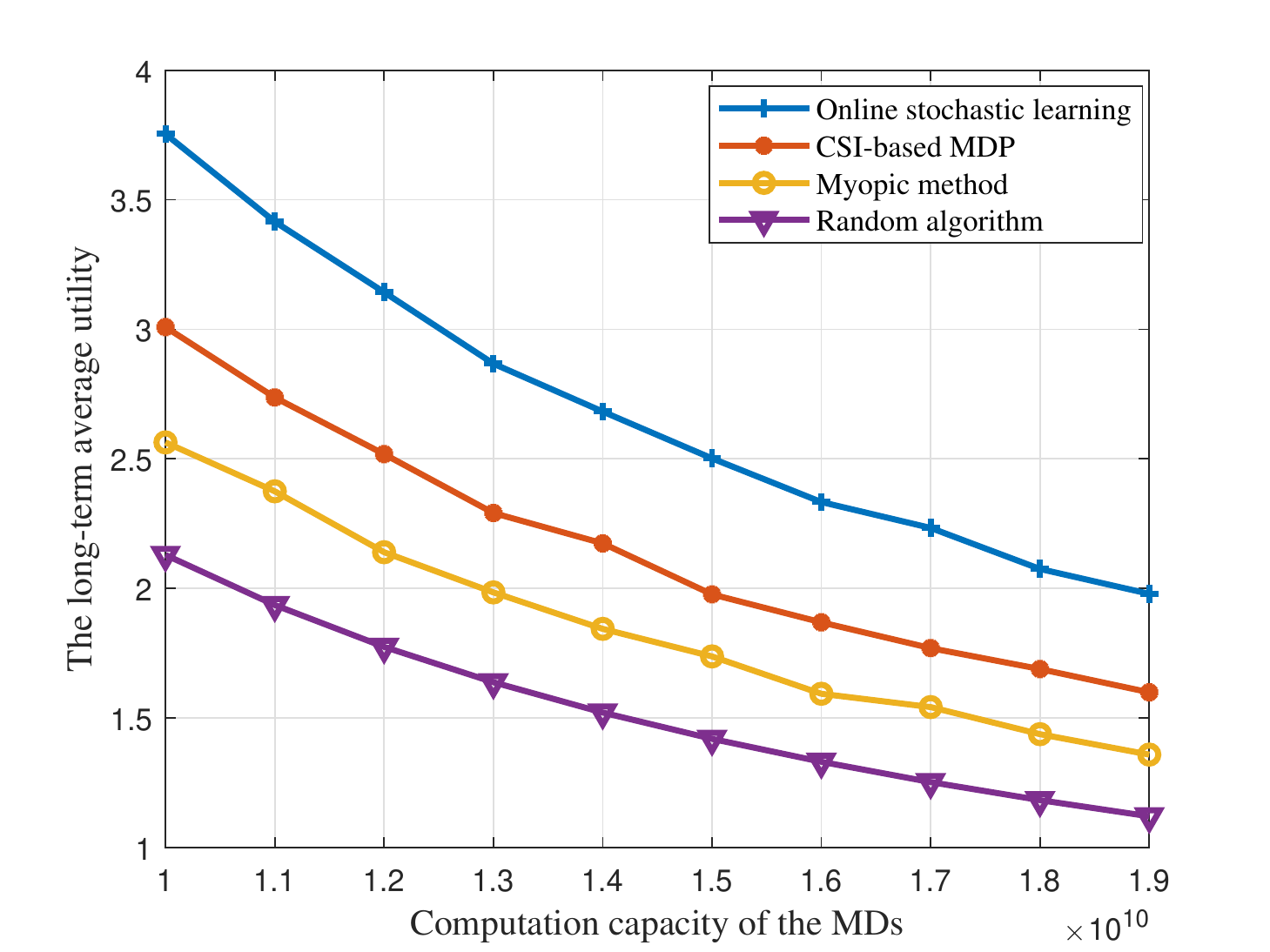}
\caption{The long-term average utility(MB) $\mathcal{U}$ v.s. the computation capacity $f$(Hz) of MD.}
\label{fig:fc_vs_utility}
\end{figure}

\begin{figure}
\centering
\includegraphics[height=6.5cm,width=8.5cm]{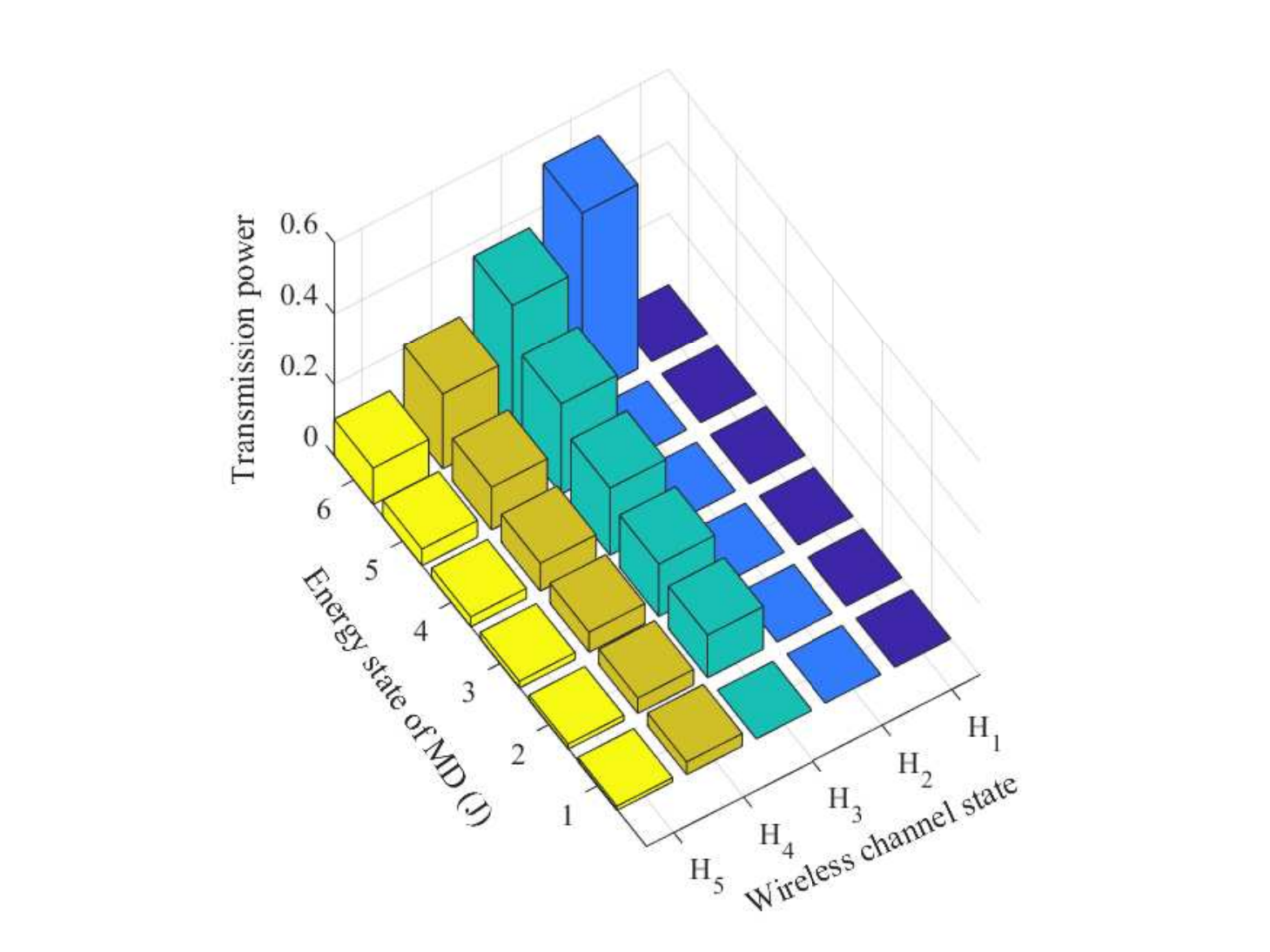}
\caption{The transmission power(W) v.s. the wireless channel state and the energy state of MD.}
\label{fig:N_vs_utility}
\end{figure}

Fig. \ref{fig:N_vs_utility} describes the relationship among the wireless channel state, the energy state of MD and the transmission power policy.
In our simulation settings, $H_1$ represents the worst channel state, while $H_5$ represents the best channel state in our system. From the figure, we can see that the MD
avoids data transmission to save battery energy when the MD is in a very poor channel state ($H_1$). In addition, we find that for a given channel state the transmission power monotonously increases with the energy state of the MD.

\begin{figure}
\centering
\includegraphics[height=6.5cm,width=8.5cm]{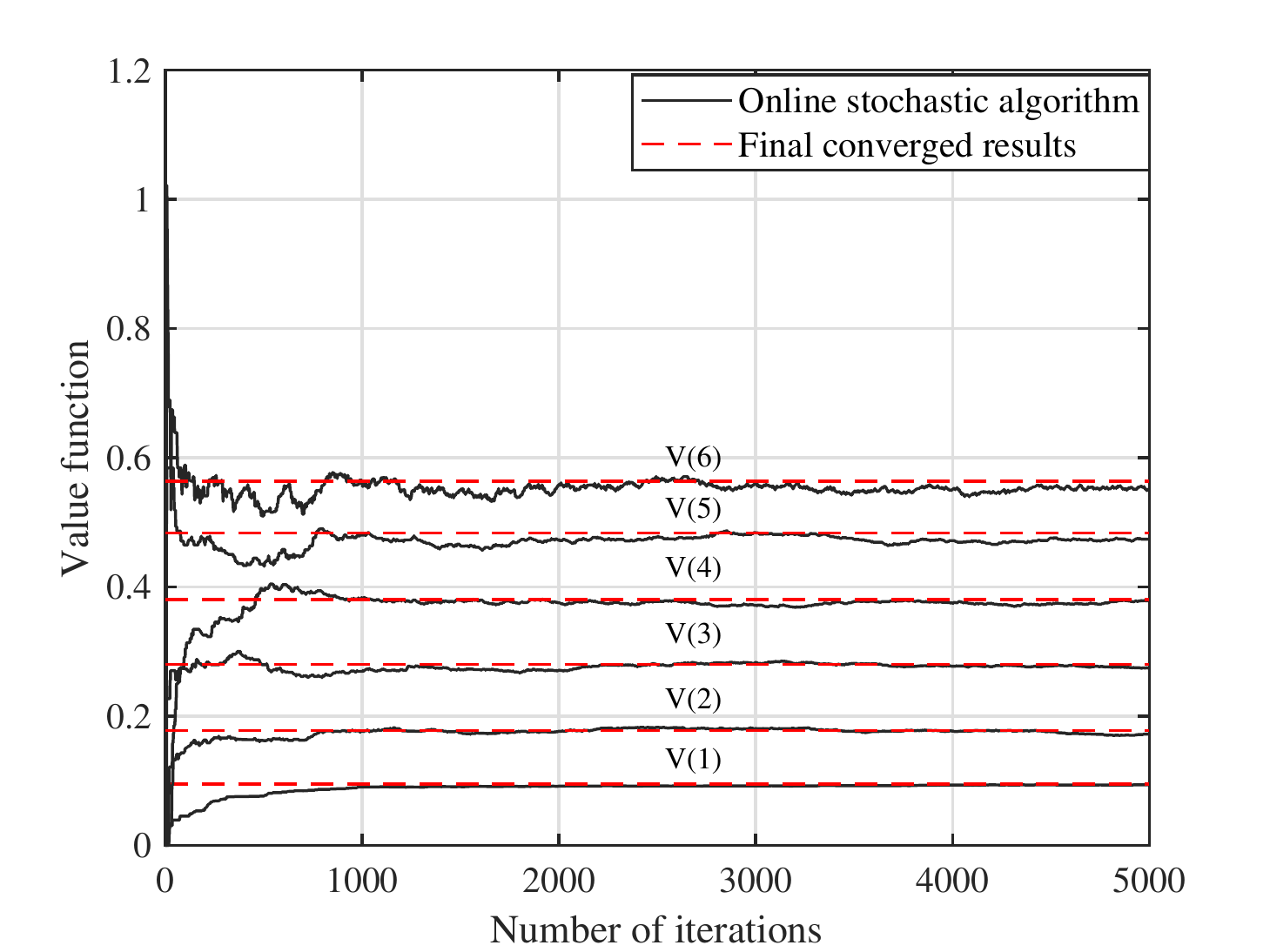}
\caption{Convergence property of the proposed online stochastic learning algorithm.}
\label{fig:value_fig}
\end{figure}

Fig. \ref{fig:value_fig} illustrates the convergence property of the proposed distributed online learning algorithm. It can be seen that the online stochastic algorithm
converges quite fast and after 1500 iterations, the values are close to the final converged results. Moreover, it is clear that the value functions calculated online quickly approach
the the final converged results when the number of iterations grows.

\section{Conclusion}
\label{sec:conclusion}
In this paper, we study a constrained MDP problem of FL with a MEC sever, where each MDs send local model updates trained on their local sensitive data  iteratively to
the edge server, and the edge server aggregates the parameters from MDs and broadcasts the aggregated parameters to MDs. We first model
the resource scheduling problem in the synchronous FL process as a constrained MDP problem, and we use the size of the training samples as the performance of FL for analysis. Due to the coupling between iterations and the complexity of the state-action space, we cannot directly solve the constrained MDP problem. Thus, we analyze the problem by equivalent Bellman equations and use approximate MDP and stochastic learning methods to simplify the constrained MDP problem so as to approximate the state value function.
Then we design static algorithm to obtain the static policy for each MD based the approximate state value function. Finally, we provide theoretical analysis for the convergence of the online stochastic learning algorithm. The simulation results show that the performance of the stochastic leaning is better than various benchmark schemes.

\ifCLASSOPTIONcaptionsoff
  \newpage
\fi



%
\bibliographystyle{IEEEtran}
\bibliography{An}

\begin{thebibliography}{10}
\providecommand{\url}[1]{#1}
\csname url@samestyle\endcsname
\providecommand{\newblock}{\relax}
\providecommand{\bibinfo}[2]{#2}
\providecommand{\BIBentrySTDinterwordspacing}{\spaceskip=0pt\relax}
\providecommand{\BIBentryALTinterwordstretchfactor}{4}
\providecommand{\BIBentryALTinterwordspacing}{\spaceskip=\fontdimen2\font plus
\BIBentryALTinterwordstretchfactor\fontdimen3\font minus
  \fontdimen4\font\relax}
\providecommand{\BIBforeignlanguage}[2]{{%
\expandafter\ifx\csname l@#1\endcsname\relax
\typeout{** WARNING: IEEEtran.bst: No hyphenation pattern has been}%
\typeout{** loaded for the language `#1'. Using the pattern for}%
\typeout{** the default language instead.}%
\else
\language=\csname l@#1\endcsname
\fi
#2}}
\providecommand{\BIBdecl}{\relax}
\BIBdecl

\bibitem{lecun2015deep}
Y.~LeCun, Y.~Bengio, and G.~Hinton, ``Deep learning,'' \emph{nature}, vol. 521,
  no. 7553, pp. 436--444, 2015.

\bibitem{mao2018deep}
Q.~Mao, F.~Hu, and Q.~Hao, ``Deep learning for intelligent wireless networks: A
  comprehensive survey,'' \emph{IEEE Communications Surveys \& Tutorials},
  vol.~20, no.~4, pp. 2595--2621, 2018.

\bibitem{wei2020federated}
K.~Wei, J.~Li, M.~Ding, C.~Ma, H.~H. Yang, F.~Farokhi, S.~Jin, T.~Q.~S. Quek,
  and H.~V. Poor, ``Federated learning with differential privacy: Algorithms
  and performance analysis,'' \emph{IEEE Transactions on Information Forensics
  and Security}, vol.~15, pp. 3454--3469, 2020.

\bibitem{wei2021user}
K.~Wei, J.~Li, M.~Ding, C.~Ma, H.~Su, B.~Zhang, and H.~V. Poor, ``User-level
  privacy-preserving federated learning: Analysis and performance
  optimization,'' \emph{IEEE Transactions on Mobile Computing}, pp. 1--1, 2021.

\bibitem{ma2020on}
C.~Ma, J.~Li, M.~Ding, H.~H. Yang, F.~Shu, T.~Q.~S. Quek, and H.~V. Poor, ``On
  safeguarding privacy and security in the framework of federated learning,''
  \emph{IEEE Network}, vol.~34, no.~4, pp. 242--248, 2020.

\bibitem{wang2019adaptive}
S.~Wang, T.~Tuor, T.~Salonidis, K.~K. Leung, C.~Makaya, T.~He, and K.~Chan,
  ``Adaptive federated learning in resource constrained edge computing
  systems,'' \emph{IEEE Journal on Selected Areas in Communications}, vol.~37,
  no.~6, pp. 1205--1221, 2019.

\bibitem{kim2019blockchained}
H.~Kim, J.~Park, M.~Bennis, and S.-L. Kim, ``Blockchained on-device federated
  learning,'' \emph{IEEE Communications Letters}, 2019.

\bibitem{mcmahan2016communication}
H.~B. McMahan, E.~Moore, D.~Ramage, S.~Hampson \emph{et~al.},
  ``Communication-efficient learning of deep networks from decentralized
  data,'' \emph{arXiv preprint arXiv:1602.05629}, 2016.

\bibitem{li2014communication}
M.~Li, D.~G. Andersen, A.~J. Smola, and K.~Yu, ``Communication efficient
  distributed machine learning with the parameter server,'' in \emph{Advances
  in Neural Information Processing Systems}, 2014, pp. 19--27.

\bibitem{alistarh2017qsgd}
D.~Alistarh, D.~Grubic, J.~Li, R.~Tomioka, and M.~Vojnovic, ``Qsgd:
  Communication-efficient sgd via gradient quantization and encoding,'' in
  \emph{Advances in Neural Information Processing Systems}, 2017, pp.
  1709--1720.

\bibitem{lin2017deep}
Y.~Lin, S.~Han, H.~Mao, Y.~Wang, and W.~J. Dally, ``Deep gradient compression:
  Reducing the communication bandwidth for distributed training,'' \emph{arXiv
  preprint arXiv:1712.01887}, 2017.

\bibitem{sattler2019sparse}
F.~Sattler, S.~Wiedemann, K.-R. M{\"u}ller, and W.~Samek, ``Sparse binary
  compression: Towards distributed deep learning with minimal communication,''
  in \emph{2019 International Joint Conference on Neural Networks
  (IJCNN)}.\hskip 1em plus 0.5em minus 0.4em\relax IEEE, 2019, pp. 1--8.

\bibitem{anh2019efficient}
T.~T. Anh, N.~C. Luong, D.~Niyato, D.~I. Kim, and L.-C. Wang, ``Efficient
  training management for mobile crowd-machine learning: A deep reinforcement
  learning approach,'' \emph{IEEE Wireless Communications Letters}, vol.~8,
  no.~5, pp. 1345--1348, 2019.

\bibitem{tran2019federated}
N.~H. Tran, W.~Bao, A.~Zomaya, N.~M. NH, and C.~S. Hong, ``Federated learning
  over wireless networks: Optimization model design and analysis,'' in
  \emph{IEEE INFOCOM 2019-IEEE Conference on Computer Communications}.\hskip
  1em plus 0.5em minus 0.4em\relax IEEE, 2019, pp. 1387--1395.

\bibitem{mohammad2019adaptive}
U.~Mohammad and S.~Sorour, ``Adaptive task allocation for asynchronous
  federated mobile edge learning,'' \emph{arXiv preprint arXiv:1905.01656},
  2019.

\bibitem{mao2016grid}
Y.~{Mao}, J.~{Zhang}, and K.~B. {Letaief}, ``Grid energy consumption and qos
  tradeoff in hybrid energy supply wireless networks,'' \emph{IEEE Transactions
  on Wireless Communications}, vol.~15, no.~5, pp. 3573--3586, 2016.

\bibitem{mao2015a}
Y.~Mao, J.~Zhang, and K.~B. Letaief, ``A lyapunov optimization approach for
  green cellular networks with hybrid energy supplies,'' \emph{IEEE Journal on
  Selected Areas in Communications}, vol.~33, no.~12, pp. 2463--2477, 2015.

\bibitem{yang2020FL}
K.~{Yang}, T.~{Jiang}, Y.~{Shi}, and Z.~{Ding}, ``Federated learning via
  over-the-air computation,'' \emph{IEEE Transactions on Wireless
  Communications}, vol.~19, no.~3, 2020.

\bibitem{wu2021multi}
X.~Wu, J.~Li, M.~Xiao, P.~C. Ching, and H.~Vincent~Poor, ``Multi-agent
  reinforcement learning for cooperative coded caching via homotopy
  optimization,'' \emph{IEEE Transactions on Wireless Communications}, pp.
  1--1, 2021.

\bibitem{wu2019dynamic}
P.~Wu, J.~Li, L.~Shi, M.~Ding, K.~Cai, and F.~Yang, ``Dynamic content update
  for wireless edge caching via deep reinforcement learning,'' \emph{IEEE
  Communications Letters}, vol.~23, no.~10, pp. 1773--1777, 2019.

\bibitem{van2018quality}
D.~Van~Le and C.-K. Tham, ``Quality of service aware computation offloading in
  an ad-hoc mobile cloud,'' \emph{IEEE Transactions on Vehicular Technology},
  vol.~67, no.~9, pp. 8890--8904, 2018.

\bibitem{zhao2019delay}
X.~Zhao, W.~Chen, J.~Lee, and N.~B. Shroff, ``Delay-optimal and
  energy-efficient communications with markovian arrivals,'' \emph{IEEE
  Transactions on Communications}, 2019.

\bibitem{Biason2018adecentralized}
A.~{Biason}, S.~{Dey}, and M.~{Zorzi}, ``A decentralized optimization framework
  for energy harvesting devices,'' \emph{IEEE Transactions on Mobile
  Computing}, vol.~17, no.~11, pp. 2483--2496, 2018.

\bibitem{Knorn2015distortion}
S.~{Knorn}, S.~{Dey}, A.~{Ahl¨¦n}, and D.~E. {Quevedo}, ``Distortion
  minimization in multi-sensor estimation using energy harvesting and energy
  sharing,'' \emph{IEEE Transactions on Signal Processing}, vol.~63, no.~11,
  pp. 2848--2863, 2015.

\bibitem{Sharma2020acceleated}
N.~{Sharma}, N.~{Mastronarde}, and J.~{Chakareski}, ``Accelerated
  structure-aware reinforcement learning for delay-sensitive energy harvesting
  wireless sensors,'' \emph{IEEE Transactions on Signal Processing}, vol.~68,
  pp. 1409--1424, 2020.

\bibitem{cui2010distributive}
Y.~{Cui} and V.~K.~N. {Lau}, ``Distributive stochastic learning for
  delay-optimal ofdma power and subband allocation,'' \emph{IEEE Transactions
  on Signal Processing}, vol.~58, no.~9, pp. 4848--4858, 2010.

\bibitem{zhao2020delay}
X.~{Zhao}, W.~{Chen}, J.~{Lee}, and N.~B. {Shroff}, ``Delay-optimal and
  energy-efficient communications with markovian arrivals,'' \emph{IEEE
  Transactions on Communications}, vol.~68, no.~3, pp. 1508--1523, 2020.

\bibitem{cui2012survey}
Y.~{Cui}, V.~K.~N. {Lau}, R.~{Wang}, H.~{Huang}, and S.~{Zhang}, ``A survey on
  delay-aware resource control for wireless systems¡ªlarge deviation theory,
  stochastic lyapunov drift, and distributed stochastic learning,'' \emph{IEEE
  Transactions on Information Theory}, vol.~58, no.~3, pp. 1677--1701, 2012.

\bibitem{zhao2019non}
X.~{Zhao} and W.~{Chen}, ``Non-orthogonal multiple access for delay-sensitive
  communications: A cross-layer approach,'' \emph{IEEE Transactions on
  Communications}, vol.~67, no.~7, pp. 5053--5068, 2019.

\bibitem{ku2015onenergy}
M.~{Ku}, W.~{Li}, Y.~{Chen}, and K.~J. {Ray Liu}, ``On energy harvesting gain
  and diversity analysis in cooperative communications,'' \emph{IEEE Journal on
  Selected Areas in Communications}, vol.~33, no.~12, pp. 2641--2657, 2015.

\bibitem{li2014efficient}
M.~Li, T.~Zhang, Y.~Chen, and A.~J. Smola, ``Efficient mini-batch training for
  stochastic optimization,'' in \emph{Proceedings of the 20th ACM SIGKDD
  international conference on Knowledge discovery and data mining}, 2014, pp.
  661--670.

\bibitem{Sudevalayam2011energy}
S.~{Sudevalayam} and P.~{Kulkarni}, ``Energy harvesting sensor nodes: Survey
  and implications,'' \emph{IEEE Communications Surveys Tutorials}, vol.~13,
  no.~3, pp. 443--461, 2011.

\bibitem{Lei2016Delay1}
L.~{Lei}, Y.~{Kuang}, N.~{Cheng}, X.~S. {Shen}, Z.~{Zhong}, and C.~{Lin},
  ``Delay-optimal dynamic mode selection and resource allocation in
  device-to-device communications¡ªpart i: Optimal policy,'' \emph{IEEE
  Transactions on Vehicular Technology}, vol.~65, no.~5, pp. 3474--3490, 2016.

\bibitem{Bettesh_2006}
I.~Bettesh and S.~Shamai, ``Optimal power and rate control for minimal average
  delay: The single-user case,'' \emph{IEEE Transactions on Information
  Theory}, vol.~52, no.~9, pp. 4115--4141, Sep. 2006.

\bibitem{Wang_Rui_2013}
R.~{Wang} and V.~K.~N. {Lau}, ``Delay-aware two-hop cooperative relay
  communications via approximate mdp and stochastic learning,'' \emph{IEEE
  Transactions on Information Theory}, vol.~59, no.~11, pp. 7645--7670, Nov.
  2013.

\bibitem{Borkar_1997}
V.~S. {Borkar}, ``Stochastic approximation with two time scales,'' \emph{Syst.
  Control Lett.}, vol.~29, pp. 291--294, 1997.

\bibitem{Borkar_2005}
V.~S. {Borkar}, ``An actor-critic algorithm for constrained markov decision
  processes,'' \emph{Syst. Control Lett.}, vol.~54, no.~3, pp. 207--213, Mar.
  2005.

\bibitem{ljung2012stochastic}
H.~J. Kushner and G.~G. Yin, \emph{Stochastic approximation and optimization of
  random systems}, 2nd~ed.\hskip 1em plus 0.5em minus 0.4em\relax New York, NY,
  USA: Springer-Verlag, 2003.

\bibitem{Lei_2016}
L.~{Lei}, Y.~{Kuang}, N.~{Cheng}, X.~{Shen}, Z.~{Zhong}, and C.~{Lin},
  ``Delay-optimal dynamic mode selection and resource allocation in
  device-to-device communications¡ªpart ii: Practical algorithm,'' \emph{IEEE
  Transactions on Vehicular Technology}, vol.~65, no.~5, pp. 3491--3505, May.
  2016.

\end{thebibliography}

%








\end{document}